\documentclass[12pt,draft,onecolumn]{IEEEtran}
\usepackage{algorithm,epsfig,cite,stfloats,bm,amssymb,amsmath,color,subfigure,
graphics,extpfeil,amsthm}
\usepackage{algorithmicx}

\newtheorem{pro}{Proposition}
\newtheorem{coro}{Corollary}

\long\def\symbolfootnote[#1]#2{\begingroup
\def\thefootnote{\fnsymbol{footnote}}
\footnote[#1]{#2}\endgroup}
\psfull

\begin{document}

% Title.
% ------
\title{Interconnection Strategies for Self-Calibration\\
of Large Scale Antenna Arrays}

\author{\normalsize{Hanyu Zhu, Fuqian Yang, Zhaowei Zhu, and Xiliang Luo$^*$}

%\thanks{This work was supported in part by the startup fund from ShanghaiTech
%University under grant no. F-0203-14-008.}

\thanks{Hanyu Zhu, Fuqian Yang, Zhaowei Zhu, and Xiliang Luo ({\tt corresponding author})
are with the School of Information Science
and Technology, ShanghaiTech University, Shanghai 201210, China
(e-mail: \{zhuhy,yangfq,zhuzhw,luoxl\}@shanghaitech.edu.cn).}
}

% make the title area
{ This paper is not submitted to IEEE Transactions on Signal Processing anymore.
Since the assumptions made in this paper, e.g. "equal calibration channel magnitudes", "equal magnitudes
in the transmit and receive RF gains", and "known calibration channels", were deemed too strong,
we are improving the results further by relaxing all the constraints and will submit updated results later. 

Thanks for reading our pre-print, all your comments are welcome.}
\maketitle

% \markboth{{
% Not be submitted to IEEE Transactions on Signal Processing anymore. The further improved work based on this paper is being submitted to  IEEE Transactions on Vehicular Technology}}{}
\renewcommand{\thepage}{}

% \hspace*{3.0cm}
% \parbox{\textwidth}{
% {
% \begin{tabular}{rl}
% % {\bf Suggested EDICS:}& SPC-COMP, SAM-CALB, SPC-PERF\\
% %{\bf Special Issue:} & {Hybrid Analog - Digital Signal Processing for}\\
% %{}& {Hardware-Efficient Large Scale Antenna Arrays}\\
% %{} & Communication Systems\\
% %{\small\bf ID Number:} & {\tt TW-Jul-15-0930}\\
% % {\bf Original Submission:} & {October 11, 2017} \\
% %{\small\bf 1st Revision:} & {\tt November 4, 2015}\\
% %{\small\bf 2nd Revision:} & {\tt January 27, 2016}\\
% %{\small\bf Accepted:} & {\tt March 8, 2016}\\
% %{\small\bf Editor:} & {\tt Dr. Jun Zhang}\\
% %{\small\bf Email:} & {\tt eejzhang@ust.hk}\\
% \end{tabular}}}

\vspace{1.5cm}
\begin{IEEEkeywords}
Calibration, self-calibration, TDD, reciprocity, Cramer-Rao Lower Bound, CRLB,
massive MIMO.
\end{IEEEkeywords}

% \newpage
\pagenumbering{arabic}\setcounter{page}{1} \markboth{{}}{}

\begin{abstract}
In time-division duplexing (TDD) systems, massive multiple-input multiple-output
(MIMO) relies on the channel reciprocity to obtain the downlink (DL) channel
state information (CSI) with the uplink (UL) CSI.
In practice, the mismatches in the radio frequency (RF) analog circuits among
different antennas at the base station (BS) break the end-to-end UL and DL channel
reciprocity. Antenna calibration is necessary to avoid the severe performance
degradation with massive MIMO. Many calibration schemes are available to
compensate the RF gain mismatches and restore the channel reciprocity in
TDD massive MIMO systems. In this paper, we focus on the internal self-calibration
scheme where different BS antennas are interconnected via hardware transmission lines.
First, we study the resulting calibration performance for an arbitrary
interconnection strategy. Next, we obtain closed-form Cramer-Rao lower
bound (CRLB) expressions for each interconnection strategy at the BS with only
$(M-1)$ transmission lines and $M$ denotes the total number of BS antennas.
Basing on the derived results, we further prove that the star interconnection
strategy is optimal for internal self-calibration due to its lowest CRLB.
In addition, we also put forward efficient recursive algorithms to derive
the corresponding maximum-likelihood (ML) estimates of all the calibration
coefficients. Numerical simulation results are also included to corroborate
our theoretical analyses and results.
\end{abstract}

\section{Introduction}\label{SecIntr}

\subsection{Massive MIMO Calibration}

In massive multiple-input multiple-output (MIMO), a large number of antennas
are installed at the base station (BS) to enhance the system spectral efficiencies
\cite{larsson14commag,lu14jstsp,resek13spmag}.
In frequency-division duplexing (FDD) systems, mobile stations (MSs) need to
feed back the downlink (DL) channel state information (CSI) to the BS in the
uplink (UL) \cite{lu14jstsp}. The consumed feedback overhead becomes
overwhelming in massive MIMO. To avoid the need to feed back the DL CSI,
time-division duplexing (TDD) is typically assumed where the channel reciprocity
can be utilized to infer the DL CSI with the UL CSI at the BS \cite{Smith2004}.
However, in practice, the transmit and receive branches are composed of
totally different analog circuits. Accordingly, the radio-frequency (RF) gain
of the transmit chain is different from that of the receive chain at the
baseband \cite{LuoMaMIMO}.
These RF gain mismatches destroy the end-to-end TDD channel reciprocity and lead
to severe performance degradation in massive MIMO systems \cite{LuoMaMIMO,wei2016twc,ZhangHardMismatch}.  Careful calibration is thus
required to compensate those RF gain mismatches at the RF front ends (FEs)
to restore the channel reciprocity. Furthermore, for some applications,
e.g. direction of arrival (DoA) estimation, accurate knowledge about the RF gains
is also required at the BS \cite{Ng1996,Zhao2016}. Thus antenna calibration is
critical to enable efficient TDD massive MIMO.

There are two main categories of calibration schemes to compensate the RF gain mismatches.
One is the ``relative calibration'' and the other one is the ``full calibration''.
The relative calibration was proposed to only restore the end-to-end UL and DL channel
reciprocity without addressing the absolute phase or amplitude coherence \cite{ShepardArgos}. On the other hand, the full calibration provides full absolute
phase and amplitude coherence between transmitters and receivers \cite{BenzinInter2017}.

To realize either the relative calibration or the full calibration,
either the ``Self-Calibration'' scheme
\cite{Nishimori2001,Liu2006,ShepardArgos,wei2016twc,Vieira2017Proposal,BenzinInter2017}
or the ``Over-The-Air (OTA)'' calibration scheme
\cite{Kalten10futnet,Shi2011,Rogalin2014,LuoCS} can be applied.
By utilizing hardware interconnections with transmission lines
\cite{Petermann2013,Nishimori2001,Liu2006,BenzinInter2017} or exploiting
the mutual coupling effects \cite{ShepardArgos,wei2016twc,Vieira2017Proposal},
the self-calibration scheme can be performed by the BS only without asking helps
from the served MSs or other antenna arrays.
Although the classical self-calibration scheme relying on hardware connection
needs extra costly analog switches and attenuators to wire all the antenna ports
together, it exhibits higher robustness and reliability in calibrating a
large scale antenna array. This is due to the fact that there are no undesired
effects, i.e. interference or reflections, that are picked up during the
calibration phase \cite{BenzinInter2017}.
The OTA calibration scheme is achieved with the help of the assisting MSs or
other antenna arrays \cite{Rogalin2014}. In massive MIMO, the OTA calibration
usually requires a significant amount of CSI feedback from the MSs \cite{LuoCS}.

\subsection{Our Work and Contributions}

The authors in \cite{BenzinInter2017} compared two interconnection strategies,
i.e. the star interconnection and the daisy chain interconnection,
for the internal self-calibration of a large scale antenna array.
However, they did not provide any theoretical analyses.
To the best of the authors' knowledge, there are few literatures addressing
the optimal interconnection strategy to connect the antennas at the BS
with transmission lines for internal self-calibration.
In this paper, we investigate this interesting and fundamental problem and
expect our results can serve as the design guidelines for massive MIMO systems.
In particular, our main technical contributions can be summarized as follows.

\begin{enumerate}
\item We obtain the closed-form expressions for the Cramer-Rao lower bounds
(CRLBs) for the calibration coefficients when an arbitrary effective interconnection
strategy using $(M-1)$ transmission lines is implemented at the BS for internal
self-calibration. The derived expressions reveal how an interconnection strategy
affects the calibration performance. In particular, we show the CRLB depends on
the number of antennas along the shortest interconnection path between one antenna
and the reference antenna;

\item For the first time, we prove that the star interconnection strategy is
optimal in the sense that it exhibits the lowest CRLB for each calibration
coefficient to be estimated. Our results can guide the designs of massive MIMO;

\item For both full calibration and relative calibration, with an arbitrary
effective interconnection strategy implemented by the BS for internal self-calibration,
we put forward efficient recursive algorithms with low complexity to derive
the maximum-likelihood (ML) estimates of the unknown calibration coefficients.
\end{enumerate}

\subsection{Paper Organization and Notations}

The rest of the paper is organized as follows.
Section \ref{SecSysModel} provides some calibration preliminaries
and gives the system model.
Section \ref{SecProCRLB} analyzes the CRLBs for the calibration coefficients
with an arbitrary interconnection strategy.
Section \ref{SecOpt} derives the closed-form CRLB expressions for one
effective interconnection strategy with $(M-1)$ interconnections
and demonstrates that the optimal interconnection strategy is the star interconnection.
In Section \ref{SecCaliMethod}, we provide recursive algorithms to derive
the ML estimates of the calibration coefficients.
Numerical results are provided in Section \ref{Secnum} and
Section \ref{SecConc} concludes the paper.

{\it Notations}:
The imaginary unit is denoted by $\jmath$.
${\sf{Diag}}\{\cdot\}$ denotes the diagonal matrix with the diagonal elements defined inside the curly brackets.
Notations ${\sf E}[\cdot]$, ${\sf Tr}\{\cdot\}$, $(\cdot)^{T}$, $(\cdot)^H$, $(\cdot)^*$, and $\vert\mathcal{C}\vert$ stand for expectation, matrix trace, transpose operation, Hermitian operation, conjugate operation, and the cardinality of the set $\mathcal{C}$, respectively.
$\mathcal{A}\setminus \mathcal{B}$ (or $\mathcal{A}-\mathcal{B}$) means the relative complement of the set $\mathcal{B}$ in the set $\mathcal{A}$.
$\Re\{\cdot\}$ and $\Im\{\cdot\}$ denote the real part and imaginary part of the argument.
Notations $\bm I_M$ ($\bm O_M$) represents the $M\times M$ identity matrix (all zero matrix), $[\bm A]_{p,q}$ denotes the $(p,q)$-th entry of matrix $\bm A$, and
$\bm L_{i,j}(l)$ denotes the elementary matrix which is the identity matrix but with an $l$ in the ($i,j$) position.

\section{System Model and Preliminaries } \label{SecSysModel}

\subsection{TDD Reciprocity Calibration}

A TDD massive MIMO system relies on the reciprocity between the UL and DL
channels to avoid the need to ask the served MSs to feed back the DL CSIs
in the UL as required by an FDD system. In particular, with the help
of the UL pilots from the served MSs, the BS can acquire the UL channels
and the BS can then design appropriate beamforming vectors with the
UL channels due to the TDD reciprocity.
In practice, the end-to-end channels also include the transceiver analog RF
circuits. Due to the fact that the transmit analog branch consists of
different RF circuits from the receive branch, the end-to-end channel
reciprocity is broken even though the physical channel between the antennas
excluding the transmit and receive circuits are still reciprocal
\cite{LuoMaMIMO}. Next we provide more preliminaries on this point.

\begin{figure}[t]
\centering
\epsfig{file=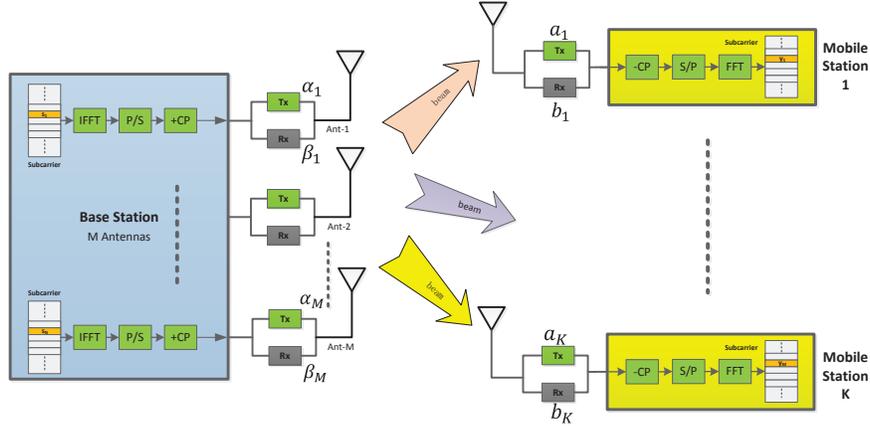,width=0.7\linewidth}
\caption{Massive MIMO system illustrating different transmit and receive
circuits break the TDD UL and DL channel reciprocity.}
\label{fig:mMIMOSys}
\end{figure}

As illustrated in Fig. \ref{fig:mMIMOSys}, we consider a large scale
multiuser TDD MIMO system with an $M$-antenna BS and $K$ single-antenna MSs.
With OFDM transmission \cite{BolcskeiOFDM}, over one particular subcarrier
in the DL, the received signals at the $K$ MSs become
\begin{equation}\label{DLsignal}
  \bm y_D=\bm H_{DL} \bm s_D+ \bm z_D,
\end{equation}
where $\bm y_D$ is a $K\times 1$ vector,
$\bm H_{DL}\in \mathbb{C}^{K\times M}$ represents the end-to-end DL channel,
$\bm s_D\in \mathbb{C}^{M\times 1}$ represents the precoded
data vector, and $\bm z_D$ denotes the receiver noise. Similarly,
In the UL, the received signals at the BS can be expressed as
\begin{equation}\label{ULsignal}
  \bm y_U=\bm H_{UL} \bm s_U+ \bm z_U,
\end{equation}
where $\bm y_U$ is an $M\times 1$ vector,
$\bm H_{UL}\in \mathbb{C}^{M\times K}$ represents the end-to-end UL channel,
$\bm s_U\in \mathbb{C}^{K\times 1}$ represents the transmitted data symbols
from the $K$ MSs, and $\bm z_U$ denotes the receiver noise at the BS.
As illustrated in Fig. \ref{fig:mMIMOSys}, we let
${\{\alpha_m, \beta_m\}}_{m=1}^M$ denote the complex-valued transmit
and receive RF gains of the antennas at the BS.
Meanwhile, we use ${\{a_m, b_m\}}_{m=1}^M$ to denote the corresponding
gains of MSs' antennas. Similar to
\cite{LuoMaMIMO,Nishimori2014,Kalten10futnet,Rogalin2014}, the
end-to-end DL and UL channel matrices can be expressed as
\begin{equation}\label{E2Echannel}
  \begin{split}
    \bm H_{DL}  &=\bm R_{MS}\bm H_{PHY}\bm T_{BS},\\
    \bm H_{UL}^T&=\bm T_{MS}\bm H_{PHY}\bm R_{BS},
  \end{split}
\end{equation}
where $\bm R_{MS}:={\sf Diag}\{b_1, b_2, \ldots, b_K\}$,
$\bm T_{MS}:={\sf Diag}\{a_1, a_2, \ldots, a_K\}$,
$\bm R_{BS}:={\sf Diag}\{\beta_1, \beta_2, \ldots, \beta_M\}$,
$\bm T_{BS}:={\sf Diag}\{\alpha_1, \alpha_2, \ldots, \alpha_M\}$,
and $\bm H_{PHY}$ denotes the propagating channel matrix which
is reciprocal under TDD operation.
From (\ref{E2Echannel}), we see the DL and UL channels are related as
\begin{equation}\label{ULDLRelation}
  \bm H_{DL}=\bm R_{MS}\bm T_{MS}^{-1}\bm H_{UL}^T\bm R_{BS}^{-1}\bm T_{BS}.
\end{equation}
It can be observed from (\ref{ULDLRelation}) that the UL channel and the
DL channel are not reciprocal when the RF gains are different, i.e.
$\bm R_{BS}^{-1}\bm T_{BS}\neq\bm I$.
For DL data detection, many works have shown the RF gain mismatches at the MSs
can be neglected and ${\{a_m, b_m\}}_{m=1}^M$ do not need to be calibrated
\cite{BenzinInter2017,wei2016twc}. On the other hand, it is critical to carry
out accurate antenna calibration at the BS. From now on, for concise
notation, we write $\bm R=\bm R_{BS}$ and $\bm T=\bm T_{BS}$.

\subsubsection{Relative vs Full Calibration}

To restore the end-to-end channel reciprocity in the presence of RF gain
mismatches at the BS, we need to adjust the gains of the transmit
or receive chains such that we have
\begin{equation}\label{ReciprocityEq}
  \bm R^{-1}\bm C_{\text{cal}}\bm T=s_c\bm I_M,
\end{equation}
where $\bm C_{\text{cal}}:=s_c\cdot{\sf Diag}\{c_1, c_2,..., c_M\}$
represents the designed calibration matrix and $s_c$ stands for the
unknown scaling coefficient.
From (\ref{ReciprocityEq}), we also see that we only need to know
the values of relative gain coefficients, i.e. $\{c_m=\beta_m/\alpha_m\}_{m=1}^M$,
to realize the end-to-end channel reciprocity.
This is called ``relative calibration'' at the BS.
For some other important applications, e.g. DoA estimation
\cite{Ng1996,Zhao2016}, the BS also needs to know the absolute phase and amplitude
coherence between all the transmit antennas and the receive antennas.
Thus ``full calibration'' at the BS should be performed to obtain all the
gain coefficients, i.e. $\{s_{\alpha}\alpha_m, s_{\beta}\beta_m\}_{m=1}^M$.
Note we allow some unknown scaling coefficients in the estimates, i.e.
$s_{\alpha}$ and $s_{\beta}$.

\subsubsection{OTA vs Self-Calibration}

Existing antenna calibration schemes for either full calibration or
relative calibration can be put into two categories:
\begin{itemize}
  \item Self-Calibration Method
  \cite{Nishimori2001,Liu2006,ShepardArgos,wei2016twc,Vieira2017Proposal,
  BenzinInter2017,Petermann2013}:
  Utilizing hardware circuits connection \cite{Petermann2013,Nishimori2001,Liu2006,BenzinInter2017} or
  mutual coupling effects \cite{ShepardArgos,wei2016twc,Vieira2017Proposal}, the
  self-calibration method can be simply run at the BS without the need of helps from
  the served MSs;

  \item OTA Calibration Method
  \cite{Kalten10futnet,Shi2011,Rogalin2014,LuoCS}:
  The OTA method calibrates the BS antenna array with the OTA feedback from the
  served MSs or the other BSs.
\end{itemize}
Note that the OTA method works well in conventional MIMO systems. However,
the amount of required CSI feedback overhead becomes overwhelming
as the size of the antenna array to be calibrated becomes large \cite{LuoCS}.
On the other hand, the self-calibration method would require
costly analog switches and attenuators wiring all the antenna ports together.
But self-calibration scheme exhibits more robustness and reliability
in calibrating a large antenna array. There are no undesired effects,
i.e. interference or reflections, are picked up during the calibration phase
\cite{BenzinInter2017}.

In this paper, we focus on the full calibration approach, which
delivers the estimates of all the unknown RF gain coefficients
(also called calibration coefficients).
Further, we will study the interconnection strategies for the internal
self-calibration method which is run at the BS.
We assume an internal wiring network which
interconnects the transmitters and receivers of the antennas at the BS internally via
transmission lines, e.g. microstrip or stripline PCB traces
\cite{BenzinInter2017}.

\subsection{Signal Model for Internal Self-Calibration}

\begin{figure}[t]
\centering
\epsfig{file=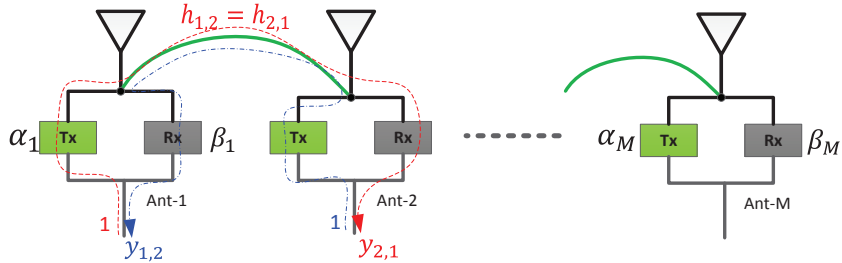,width=0.7\linewidth}
\caption{Signal model for self-calibration with internal interconnection
wiring. ($y_{1,2}$: the received calibration measurement at antenna-$1$ due to
the transmission from antenna-$2$; $y_{2,1}$: the received calibration measurement
at antenna-$2$ due to the transmission from antenna-$1$. Note
the calibration channels between antenna-$1$ and antenna-$2$ are reciprocal, i.e.
$h_{1,2}=h_{2,1}$.) }
\label{fig:sigModel}
\end{figure}

During the calibration phase, the BS antennas transmit
sounding signals over the transmission lines to obtain calibration measurements.
Let $y_{p,q}$ denote the received signal at the $p$-th antenna due to the
transmission from the $q$-th antenna. Without loss of generality, we assume
the sounding signal is $1$. We then have
\begin{equation}\label{scalarrece}
  y_{p,q}= \beta_p h_{p,q} \alpha_q+n_{p,q},
\end{equation}
where $h_{p,q}$ represents the gain of the calibration channel between the
$p$-th antenna and $q$-th antenna and $n_{p,q}$ is additive white Gaussian
noise (AWGN) with zero mean and variance $\sigma_n^2$. See also Fig. \ref{fig:sigModel}.
Note we have $h_{p,q}=0$ if there is no interconnection wiring between the $p$-th
antenna and the $q$-th antenna.
Furthermore, we have $h_{p,q}=h_{q,p}$ due to the reciprocity of the
calibration channel. By stacking all the calibration measurements in
(\ref{scalarrece}) together, we can have the following matrix form:
\begin{equation}\label{matrixrece}
\bm{Y}=\bm R \bm H \bm T+\bm{N},
\end{equation}
where $[\bm Y]_{p,q}:=y_{p,q}$,
${\bm R}:={\sf Diag}\{\beta_1,\beta_2,\cdots,\beta_M\}$,
${\bm T}:={\sf Diag}\{\alpha_1,\alpha_2,\cdots,\alpha_M\}$,
$[\bm H]_{p,q}:=h_{p,q}$, and $[\bm N]_{p,q}:=n_{p,q}$.
Note that $\bm H=\bm H^T$ since the calibration channels over
each transmission line are reciprocal, i.e. $h_{p,q}=h_{q,p}$.

In this paper, we endeavor to find the optimal interconnection strategy or
wiring at the BS such that different antennas are connected in the most efficient
way to enable the best estimates of the calibration coefficients.
To proceed with our derivations, we first make the following assumption:
\begin{itemize}
  \item{\sf AS-1}: All the transmission lines have the same length and damping,
  i.e. $h_{p,q}=h$ when the $p$-th antenna and the $q$-th antenna are
  interconnected.
\end{itemize}
In practice, when all the transmission lines have the same length,
it could become hard to handle from the hardware implementation point of view in
massive MIMO systems. For example, some particular wiring methods could lead to excessive
meandering of the transmission
lines to some antennas \cite{BenzinInter2017}.

With AS-1, the calibration signal model in (\ref{matrixrece}) can
be simplified to
\begin{equation}\label{matrixrece2}
  \bm Y=h\bm R \mathcal{\bm A} \bm T+\bm N,
\end{equation}
where the matrix $\mathcal{A}$ represents the interconnection. Specifically,
it is defined as
\begin{equation}
\mathcal{A}_{p,q} :=
\left\{\begin{array}{cc}
1, &\text{Antenna-}p,q\text{ are interconnected}\\
0, &\text{otherwise}
\end{array} \right..
\end{equation}

\section{Performance Analysis of An Arbitrary Interconnection Strategy}\label{SecProCRLB}

To restore the end-to-end channel reciprocity, we only need to
know the values of those transmit and receive RF gains subject to a
common scaling, e.g. $\{s_{\alpha}\alpha_m\}_{m=1}^M$ and
$\{s_{\beta}\beta_m\}_{m=1}^M$. In order to proceed with
our quantitative analyses, we assume there is a ``reference antenna'',
e.g. the $f$-th antenna, whose RF gains: $\alpha_f$ and $\beta_f$ are known.
The other antennas are termed ``ordinary antennas'' accordingly.
For a particular interconnection strategy, given all the measurements $\bm Y$
in (\ref{matrixrece}), we can
derive the corresponding CRLBs for those unknown calibration
coefficients, i.e. ${\left\{\alpha_m,\beta_m\right\}}_{m=1}^M\setminus
\{\alpha_{f},\beta_{f}\}$. Note these CRLBs serve as the
lower bounds for the variances of the estimation errors of all possible
unbiased estimators \cite{Kay1993}.

Note that (\ref{matrixrece}) can be rewritten in the following vector
form:
\begin{equation}\label{vecmodel}
  \bm y=\bm \mu+\bm n,
\end{equation}
where
\begin{equation}
\begin{split}
\bm y&:=[\tilde{\bm y}_{1,2}^T, \ldots, \tilde{\bm y}_{1,M}^T, \tilde{\bm y}_{2,3}^T, \ldots, \tilde{\bm y}_{2,M}^T, \ldots, \tilde{\bm y}_{M-1,M}^T]^T,\\
\tilde{\bm y}_{p,q}&:=\left[y_{p,q},y_{q,p}\right]^T,\\
{\bm{\mu}}&:=[\tilde{\bm \mu}_{1,2}^T,\ldots,\tilde{\bm \mu}_{1,M}^T,\tilde{\bm\mu}_{2,3}^T,\ldots,\tilde{\bm \mu}_{2,M}^T,\ldots,\tilde{\bm \mu}_{M-1,M}^T]^T,\\
\tilde{\bm \mu}_{p,q}&:=\left[\beta_p h_{p,q}\alpha_q, \beta_q h_{q,p} \alpha_p\right]^T,
\end{split}
\end{equation}
and $\bm n$ is the corresponding AWGN measurement noise vector.
We can define a $2(M-1)$-by-$1$ vector $\tilde{\bm{\theta}} $ as
\begin{equation}\label{Ctheta}
 \tilde{\bm{\theta}}:=[\bm \alpha^T, \bm \beta^T]^T,
\end{equation}
where
$\bm \alpha:=\left[\alpha_1,...,\alpha_{f-1},\alpha_{f+1},...,\alpha_M\right]^T$
and
$\bm \beta:=\left[\beta_1,...,\beta_{f-1},\beta_{f+1},\ldots, \beta_M\right]^T$.
With the signal model in (\ref{vecmodel}), the probability density function
(PDF) of the measurements follows the complex Gaussian form as:
\begin{equation}\label{gaussianpdf}
  p(\bm y\vert\tilde{\bm\theta})=\frac{1}{\pi^{2(M-1)}\text{det}(\bm \Sigma)}
  \exp\left\{-(\bm y-\bm \mu)^H \bm \Sigma^{-1} (\bm y-\bm \mu)\right\},
\end{equation}
where $\bm \Sigma=\sigma_n^2 \bm I$ is the covariance matrix of $\bm n$.
Define the Fisher information matrix of the complex parameter $\tilde{\bm \theta}$
in (\ref{Ctheta}) as $\bm J(\tilde{\bm{\theta}})$.
Let $\bar{\mathcal{A}}$ denote the submatrix obtained by removing the
$f$-th row and the $f$-th column from the interconnection matrix $\mathcal{A}$.
Now we can establish the following result. Detailed derivations can be
found in Appendix \ref{AppendixFIM}.

\begin{pro}\label{pro1}
Considering a BS with $M$ antennas interconnected with a strategy $\mathcal{A}$,
under AS-1, with the calibration signal model
in (\ref{matrixrece2}), we can obtain the CRLB matrix
for $\tilde{\bm \theta}$ as
\begin{equation}\label{CRLB_full}
{\sf CRLB}(\tilde{\bm{\theta}}|\mathcal{A})=
  \left(\bm J(\tilde{\bm{\theta}})\right)^{-1},
\end{equation}
where the Fisher information matrix $\bm J(\tilde{\bm{\theta}})$ is given by
\begin{equation}
\label{FishCtheta2}
\bm J(\tilde{\bm{\theta}})=
\frac{\vert h\vert^{2}}{\sigma_n^2}\cdot
\left[\begin{array}{cc}
   \bm A& \bm D^H\\
   \bm D&\bm B
\end{array} \right],
\end{equation}
with
\begin{equation}
\begin{split}
\bm D &:={\sf Diag}\left\{\bm{\beta}\right\}\cdot\bar{\mathcal{A}}\cdot
{\sf Diag}\{\bm{\alpha^H}\},\\
\bm A &:={\sf Diag}\left\{\sum\limits_{i\in\mathcal{C}_1}|\beta_i|^2, \ldots, \sum\limits_{i\in\mathcal{C}_m, m\ne f}|\beta_i|^2, \ldots,
\sum\limits_{i\in\mathcal{C}_M}|\beta_i|^2\right\},\\
\bm B &:={\sf Diag}\left\{\sum\limits_{i\in\mathcal{C}_1}|\alpha_i|^2, \ldots,
\sum\limits_{i\in\mathcal{C}_m, m\ne f}|\alpha_i|^2, \ldots,
\sum\limits_{i\in\mathcal{C}_M}|\alpha_i|^2\right\},
\end{split}
\end{equation}
and $\mathcal{C}_m$ denotes the set of the indices of the antennas
that are interconnected to the $m$-th antenna directly in this particular
interconnection strategy $\mathcal{A}$.
\end{pro}

\begin{figure}[t]
\centering
\epsfig{file=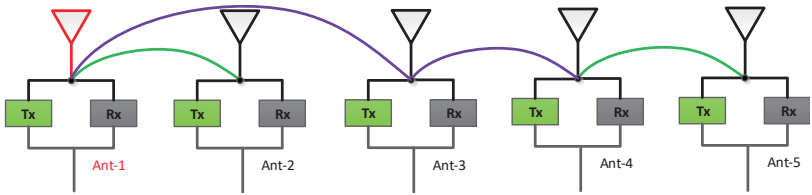,width=0.65\textwidth}
\caption{The interconnection strategy with $5$ antennas.
Antenna-$1$ is chosen as the reference antenna and the purple path
determine the calibration path of antenna-$4$.}
\label{fig:pathexample}
\end{figure}

We call the shortest interconnection path between one ordinary antenna and the
reference antenna a ``calibration path''. For example, the purple path shown
in Fig. \ref{fig:pathexample} is the calibration path of antenna-$4$.
To be able to estimate all the calibration
coefficients, the chosen interconnection strategy $\mathcal{A}$ must be
``effective'' in the sense that there must be a calibration path between each
ordinary antenna and the reference antenna.
Note that in Proposition \ref{pro1}, to ensure that the Fisher information
matrix in (\ref{FishCtheta2}) is invertible, the interconnection
strategy $\mathcal{A}$ has to be effective.

In practice, we have a total budget of $N=N_0$ transmission lines
to interconnect different antenna ports at the BS. This is due to
the cost consideration and the floor plan limitation.
Given $N_0$ transmission lines, we can find the optimal
interconnection strategy $\mathcal{A}$ by solving the following
optimization problem:
\begin{equation}
\label{OptPro}
\begin{array}{cc}
\underset{\mathcal{A}}{\rm minimize}
& {\sf Tr}\left\{{\sf CRLB}(\tilde{\bm \theta}|\mathcal{A})\right\}\\
{\rm subject\ to} & N=N_0.
\end{array}
\end{equation}

To ensure an effective interconnection strategy, we must have
at least $(M-1)$ transmission lines, i.e. we need to ensure $N_0\ge M-1$ in
(\ref{OptPro}). In general, with $N_0$ transmission lines, the total
number of effective interconnection strategies that can connect all the BS antennas together is finite. The optimization problem in (\ref{OptPro}) can be solved
by exhaust searching. However, for a large scale antenna array, it becomes
hard to handle due to the large number of antennas at the BS.
In next section, we will look further into the optimization
problem in (\ref{OptPro}) and obtain some insightful guidelines in designing
the interconnection strategy for full calibration at the BS.

\section{Optimal Interconnection Strategy with $(M-1)$ Transmission Lines}\label{SecOpt}

In this section, assuming a total budget of $(M-1)$ transmission lines to
interconnect the antennas, we further examine the CRLBs for the calibration
coefficients given by Proposition 1. To gain more insights from
(\ref{CRLB_full}), we further make the following assumption:
\begin{itemize}
  \item{\sf AS-2}: The transmit and receive RF gains exhibit equal amplitudes, i.e.
  $\vert\alpha_m\vert=a, \vert\beta_m\vert=b$, $\forall m\in[1,M]$.
\end{itemize}
AS-2 is made here mainly due to the following concern.
Constant transmit and receive amplitudes ensure identical receive
signal-to-noise ratio (SNR) in the calibration measurements at each BS antenna.
In general, the SNR in the received measurements will directly affect the
estimation performance of the calibration coefficients.
In our current study, we try to focus on the impacts of the internal
interconnection strategy.

Under AS-2, we are able to obtain closed-form expressions for the CRLBs in (\ref{CRLB_full}). Further, we
will characterize the optimal interconnection strategies for internal full
calibration and relative calibration according to the derived analytical results.

\subsection{Optimal Interconnection Strategy for Full Calibration}\label{SubsecOptConn}

From now on, we assume that we have a total number of $(M-1)$ transmission
lines to deploy at the BS. From previous discussion, we know this is the least
number of transmission lines that can ensure the interconnection strategy is
effective. In fact, we have an $M$-vertex connected graph with $(M-1)$ edges.
From Theorem $2.1.4$ in \cite{2001_GraphTheory}, we can readily draw the following
conclusion: {\it the calibration path of each ordinary antenna is unique
in every effective interconnection strategy with $(M-1)$ transmission lines}.

Under AS-2, we have $|\alpha_m|=a$ and $|\beta_m|=b$.
Then the Fisher information matrix in (\ref{FishCtheta2}) can be rewritten as
\begin{equation}\label{FishCtheta3}
\bm J(\tilde{\bm{\theta}})=\frac{|h|^{2}}{\sigma_n^2}\cdot
\left[\begin{array}{cc}
{\sf Diag}\{b^2\cdot\bm{ t}\}&\bm D^H\\
\bm D&{\sf Diag}\{a^2\cdot\bm{ t}\}
\end{array}\right],
\end{equation}
where $\bm{t}:=[t_1,...,t_{f-1},t_{f+1},..., t_M]^T$ and
$t_m=\vert\mathcal{C}_m\vert\geq1$ denotes the number of antennas that are
connected to the $m$-th antenna directly. Regarding the interconnection strategy
at the BS, we can have the following useful result. The detailed proof
of the following proposition is given in Appendix \ref{AppendixPro}.

\begin{pro}\label{pro2}
Assuming an $M$-antenna BS with one reference antenna, i.e. the $f$-th antenna, and
$(M-1)$ ordinary antennas, for an interconnection strategy $\mathcal{A}$ consuming
$(M-1)$ transmission lines, when some ordinary antennas are not interconnected
to the reference antenna, there exists one ordinary antenna which is
only connected to another ordinary antenna, i.e.
$\exists n\neq f$ and $\exists u\neq f$, such that $h_{n,u}\neq 0$ and
$h_{n,q}=0, \forall q\neq u$.
\end{pro}

\begin{figure}[t]
\centering
\epsfig{file=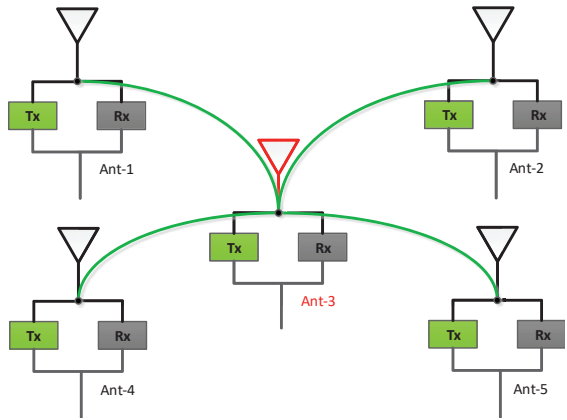,width=0.45\textwidth}
\caption{The star interconnection strategy with $5$ antennas.
Antenna-$3$ is chosen as the reference antenna and all the other $4$
ordinary antennas are interconnected to the reference antenna.}
\label{fig:Conv}
\end{figure}

Now, let's take a look at the star interconnection strategy as shown in
Fig. \ref{fig:Conv}, where all the ordinary antennas are connected to the
reference antenna. In this case, the Fisher information matrix
$\bm J(\tilde{\bm{\theta}})$ becomes diagonal since $\bar{\mathcal{A}}=\bm O$.
Specifically, the Fisher information matrix in (\ref{FishCtheta3}) becomes
\begin{equation}\label{FIMconv}
\bm J_{\text{star}}(\tilde{\bm{\theta}})=\frac{|h|^{2}}{\sigma_n^2}
\left[
  \begin{array}{cc}
    \bm A &\bm O\\
    \bm O & \bm B
  \end{array}
  \right],
\end{equation}
where
\begin{equation}
  \begin{split}
    \bm A&={\sf Diag}\left\{b^2, b^2, \ldots, b^2\right\},\\
    \bm B&={\sf Diag}\left\{a^2, a^2, \ldots, a^2\right\}.
  \end{split}
\end{equation}

If some ordinary antennas are connected to other ordinary antennas, we see
$\bar{\mathcal{A}}\ne \bm O$ and the Fisher information
matrix $\bm J(\tilde{\bm{\theta}})$ is not diagonal anymore.
Regarding this kind of interconnection strategies at the BS where some ordinary
antennas are interconnected together, based on Proposition 2, we further
establish the following useful result. See Appendix \ref{AppendiaxFimupdate}
for the proof.
\begin{pro}\label{pro3new}
Considering a BS with $M$ antennas interconnected with a strategy $\mathcal{A}^{(k)}$
where $1\le W\le M-2$ ordinary antennas are not interconnected to the reference
antenna, under AS-1 and AS-2, with the calibration signal model in
(\ref{matrixrece2}), we can obtain an updated interconnection strategy
$\mathcal{A}^{(k+1)}$ where only $(W-1)$ ordinary antennas are not interconnected
to the reference antenna directly from the strategy $\mathcal{A}^{(k)}$.
In particular, in $\mathcal{A}^{(k)}$, we can find one ordinary antenna, i.e.
the $n_k$-th antenna, which is only connected to another ordinary antenna, i.e.
the $u_k$-th antenna. Then we just disconnect the connection to the $u_k$-th antenna
and interconnect the $n_k$-th antenna to the reference antenna instead. After this
update, we have the interconnection strategy $\mathcal{A}^{(k+1)}$, where $h_{n_k,u_k}=0$
and $h_{n_k,f}\ne 0$. Furthermore, we can obtain the relationship between the Fisher
information matrices for $\mathcal{A}^{(k)}$ and $\mathcal{A}^{(k+1)}$ as
\begin{equation}\label{Fimtransk}
     \bm J^{(k+1)}(\tilde{\bm{\theta}})=\bm L^{(k)} \bm J^{(k)}(\tilde{\bm{\theta}}) \bm L'^{(k)},
\end{equation}
where
\begin{eqnarray}\label{ele_mat}
  \bm L^{(k)} := \bm L_{\bar u'_k,\bar n_k}\left(-\frac{\beta_{u} \alpha_{n}^*}{b^2}\right) \bm L_{\bar u_k,\bar n'_k}\left(-\frac{\beta_{n}^* \alpha_{u}}{a^2}\right),\\
  \bm L'^{(k)} := \bm L_{\bar n'_k,\bar u_k}\left(-\frac{\beta_{n} \alpha_{u}^*}{a^2}\right) \bm L_{\bar n_k,\bar u'_k}\left(-\frac{\beta_{u}^* \alpha_{n}}{b^2}\right),
\end{eqnarray}
$\bm L_{\cdot,\cdot}(\cdot)$ represents the elementary matrix,
and $\bm J^{(k)}(\tilde{\bm{\theta}})$ denotes the Fisher information matrix
corresponding to the interconnection strategy $\mathcal{A}^{(k)}$.
Note that $\bar n_k$ and $\bar n'_k$ denote the indices of the rows corresponding to
$\alpha_{n_k}$ and $\beta_{n_k}$ in $\tilde{\bm{\theta}}$ respectively and
$\bar u_k$ and $\bar u'_k$ denote the indices of the rows corresponding to
$\alpha_{u_k}$ and $\beta_{u_k}$ in $\tilde{\bm{\theta}}$ respectively.
\end{pro}

\begin{figure}[t]
\centering
\epsfig{file=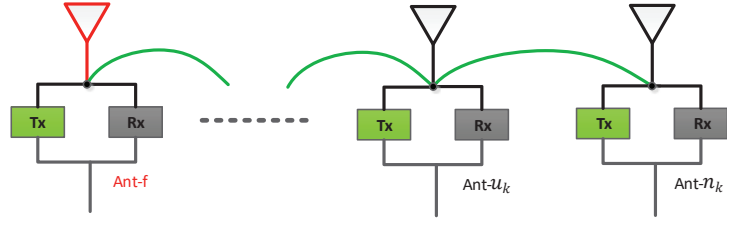,width=0.58\textwidth}
\caption{The interconnection network corresponding to the interconnection strategy
$\mathcal{A}^{(k)}$. The $f$-th antenna is chosen as the reference antenna and
the $n_k$-th antenna is only interconnected to the $u_k$-th antenna.}
\label{fig:original}
\end{figure}

\begin{figure}[t]
\centering
\epsfig{file=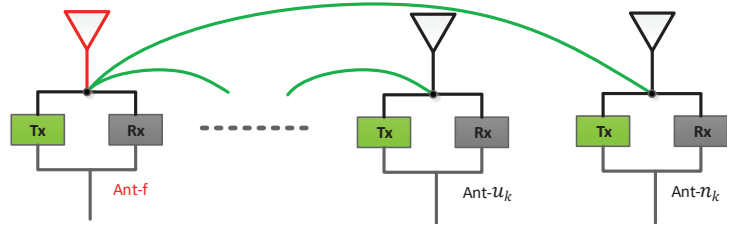,width=0.58\textwidth}
\caption{The interconnection network corresponding to the interconnection strategy
$\mathcal{A}^{(k+1)}$. The $f$-th antenna is chosen as the reference antenna and the
$n_k$-th antenna is now interconnected to antenna-$f$ after the update as
described in Proposition \ref{pro3new}.}
\label{fig:updated}
\end{figure}

In Fig. \ref{fig:original} and Fig. \ref{fig:updated}, we have illustrated the
aforementioned update in Proposition 3.
By taking inverse of both sides of (\ref{Fimtransk}), we have the relationship
between the CRLBs of the interconnection strategies $\mathcal{A}^{(k)}$ and $\mathcal{A}^{(k+1)}$, i.e.
\begin{equation}\label{CRLBk}
\left(\bm J^{(k)}(\tilde{\bm{\theta}})\right)^{-1}=\bm L'^{(k)} \left(\bm J^{(k+1)}(\tilde{\bm{\theta}})\right)^{-1} \bm L^{(k)}.
\end{equation}
Note that the CRLB matrix is updated in the reversed order
in (\ref{CRLBk}). Specifically, the matrix
$(\bm J^{(k)}(\tilde{\bm{\theta}}))^{-1}$ is updated
with the matrix $(\bm J^{(k+1)}(\tilde{\bm{\theta}}))^{-1}$.

\begin{figure}[t]
\centering
\epsfig{file=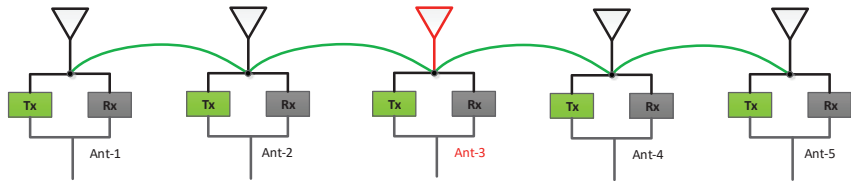,width=0.68\textwidth}
\caption{The daisy chain interconnection strategy with $5$ antennas with
antenna-$3$ being selected as the reference antenna.}
\label{fig:daisy}
\end{figure}

According to Proposition \ref{pro2} and Proposition \ref{pro3new}, given the
original interconnection strategy $\mathcal{A}^{(0)}$ where $W$ ($W\in[1,M-2]$)
ordinary antennas are not interconnected to the reference antenna, we can
obtain $\bm J^{(k)}(\tilde{\bm{\theta}})$ by applying $k$ appropriate
updates as shown in (\ref{Fimtransk}). Note that in the interconnection strategy
$\mathcal{A}^{(k)}$, there are only $(W-k)$ ordinary antennas which are not
interconnected to the reference antenna. Hence, after $W$ updates,
the strategy $\mathcal{A}^{(W)}$ becomes the star interconnection strategy
since each antenna is interconnected to the reference antenna in
$\mathcal{A}^{(W)}$. In other words, after a series of elementary
transformations, we will end up with
$\bm J^{(W)}(\tilde{\bm{\theta}})=\bm J_{\text{star}}(\tilde{\bm{\theta}})$,
which corresponds to the star interconnection strategy as shown in
Fig. \ref{fig:Conv}. Specifically, we have
\begin{equation}\label{FIMrela}
\begin{split}
\bm J_{\text{star}}(\tilde{\bm{\theta}})=
\bm J^{(W)}(\tilde{\bm{\theta}})
=&\bm L^{(W-1)}\cdots \bm L^{(0)}\cdot \bm J^{(0)}(\tilde{\bm{\theta}})\cdot \\
 &\bm L'^{(0)}
\cdots \bm L'^{(W-1)}.
\end{split}
\end{equation}
By taking inverse of both sides of (\ref{FIMrela}), we further obtain
\begin{equation}\label{CRLBrelation}
\begin{split}
\left(\bm J^{(0)}(\tilde{\bm{\theta}})\right)^{-1}
=&\bm L'^{(0)}\cdots \bm L'^{(W-1)}\cdot \left(\bm J_{\text{star}}(\tilde{\bm{\theta}})\right)^{-1} \cdot \\
& \bm L^{(W-1)}\cdots \bm L^{(0)}.
\end{split}
\end{equation}
From the result in (\ref{CRLBrelation}), we can establish the following proposition
and the proof is outlined in Appendix \ref{AppendixPro4}.
\begin{pro}\label{pro4}
Considering a BS with $M$ antennas interconnected with $(M-1)$ transmission lines,
under AS-1 and AS-2, for any kind of effective interconnection strategy,
the CRLBs for $\alpha_m$ and $\beta_m$, $\forall m\ne f$, are given by
\begin{equation}\label{CRLBpro3}
\begin{split}
    &{\sf CRLB}(\alpha_m)= \frac{(d_m+1)\sigma_n^2}{b^2\left\vert h\right\vert^2},\\
    &{\sf CRLB}(\beta_m)= \frac{(d_m+1)\sigma_n^2}{a^2\left\vert h\right\vert^2},
\end{split}
\end{equation}
where $d_m$ denotes the number of antennas along the calibration path
between the reference antenna and the $m$-th antenna in addition to
the reference antenna and the $m$-th antenna\footnote{For the interconnection
illustrated in Fig. \ref{fig:daisy}, we have
$d_1=d_5=1$ for antenna-$1$ and antenna-$5$,
$d_2=d_4=0$ for antenna-$2$ and antenna-$4$.}.
\end{pro}

From Proposition \ref{pro4}, it can be observed that the CRLBs for the unknown
parameters $\alpha_m$ and $\beta_m$ are directly determined by the calibration path
of the $m$-th antenna and the SNR in the corresponding calibration measurement.
Furthermore, Proposition \ref{pro4} provides nice closed-form results for an arbitrary
interconnection strategy. For example, when the BS implements the daisy chain
interconnection strategy \cite{BenzinInter2017} as shown in Fig. \ref{fig:daisy},
where antenna-$m$ and antenna-$(m+1)$ are interconnected, $\forall m\in[1,M-1]$,
we have $d_1=d_5=1$ and $d_2=d_4=0$. Then the corresponding CRLBs can be computed
readily according to (\ref{CRLBpro3}).

In \cite{BenzinInter2017}, the authors have shown the daisy chain interconnection
strategy would suffer from an error propagation effect assuming noisy calibration
measurements. In Proposition \ref{pro4}, we indeed show the error propagation effect
on the calibration performance for any effective interconnection strategy with
$(M-1)$ transmission lines. Specifically, the CRLBs in (\ref{CRLBpro3})
indicate that, when the calibration path of the $m$-th ordinary antenna consists of
more antennas, the calibration performance will decrease accordingly.

From Proposition \ref{pro4}, we can also easily draw the conclusion that
the minimum CRLB is obtained when $d_m=0$, $\forall m\neq f$.
Obviously, the star interconnection strategy as shown in Fig. \ref{fig:Conv}
can achieve the minimum total CRLB for the calibration coefficients, i.e.
\begin{equation}
\min_{\mathcal{A}}{\sf Tr}\left\{{\sf CRLB}(\tilde{\bm{\theta}}|\mathcal{A})\right\}
={\sf Tr}\left\{\left(\bm J_{\text{star}}(\tilde{\bm{\theta}})\right)^{-1}\right\}.
\end{equation}
Summarizing we can put forward the following important corollary.
\begin{coro}\label{coro1}
Considering a BS with $M$ antennas interconnected with $(M-1)$ transmission lines,
under AS-1 and AS-2, in order to minimize the total CRLB for all the unknown
calibration coefficients during internal self-calibration,
we should implement the star interconnection strategy.
\end{coro}

Corollary \ref{coro1} shows that the star interconnection strategy can achieve
the minimum total CRLB for full calibration during the internal
self-calibration. Thus the optimal solution of the optimization problem in
(\ref{OptPro}) is the star interconnection. Clearly, Corollary 1 can
serve as a design philosophy for internal self-calibration in massive MIMO.
In practice, compared with other interconnection strategies, the star interconnection
strategy may consume more time resources for signal exchanges and longer transmission
lines to interconnect the antennas.

\subsection{Optimal Interconnection Strategy for Relative Calibration}
\label{SubsecOptRelative}

In previous subsection, we have analyzed the optimal interconnection strategy
for full calibration at the BS. In fact, similar results can be obtained in
the case of relative calibration as well.

Let $c_m:=\beta_m/\alpha_m$, $m\in [1,M]$, denote the relative calibration
coefficients to be estimated. Assume $c_f$ is known, i.e. the $f$-th antenna
serves as the reference antenna. The CRLBs for the relative calibration
coefficients can be obtained from the CRLBs for $\tilde{\bm \theta}$ \cite{Kay1993}.
In particular, we can obtain
\begin{equation}\label{CRLBrela}
{\sf CRLB}(\bm c|\mathcal{A})=\frac{\partial \bm g(\tilde{\bm \theta})}
{\partial \tilde{\bm \theta}}{\sf CRLB}(\tilde{\bm \theta})
\frac{\partial \bm g(\tilde{\bm \theta})}{\partial \tilde{\bm \theta}}^H,
\end{equation}
where $\bm c:=[c_1,\ldots,c_{f-1},c_{f+1}, \ldots, c_M]^T$,
the $\bar{m}$-th entry of $\bm g(\tilde{\bm \theta})$ is $c_m$, i.e.
$g_{\bar m}(\tilde{\bm \theta})=c_m={\beta_m}/{\alpha_m}$,
$\forall m\in\left\{1, \ldots,f-1,f+1,\ldots, M\right\}$, and
$\bar m$ denotes the index of the element $c_m$ in $\bm{c}$.
Note ${\partial \bm g(\tilde{\bm \theta})}/{\partial \tilde{\bm \theta}}$ in
(\ref{CRLBrela}) is an $(M-1)$-by-$2(M-1)$ Jacobian matrix whose
$\bar m$-th row is ${\partial g_{\bar m}(\tilde{\bm \theta})}/
{\partial \tilde{\bm \theta}}$. It can be verified that all the other
elements are zeros except the ${\bar m}$-th element and the
$(M+{\bar m}-1)$-th element in ${\partial g_{\bar m}(\tilde{\bm \theta})}/
{\partial \tilde{\bm \theta}}$. Further, these two non-zero elements
are given by
\begin{equation}
\begin{split}
\frac{\partial  g_{\bar m}(\tilde{\bm \theta})}{\partial \alpha_m}
&=-\frac{\beta_m}{\alpha_m^2},\\
\frac{\partial  g_{\bar m}(\tilde{\bm \theta})}{\partial \beta_m}
&=\frac{1}{\alpha_m}.
\end{split}
\end{equation}
According to the results in (\ref{CRLBpro3}), the CRLB for the relative
calibration coefficients $c_m$ can be expressed as
\begin{equation}\label{CRLBc}
\begin{split}
{\sf CRLB}(c_m)&=\frac{\partial  g_{\bar m}(\tilde{\bm \theta})}{\partial \tilde{\bm \theta}}{\sf CRLB}(\tilde{\bm \theta})\frac{\partial  g_{\bar m}(\tilde{\bm \theta})}{\partial \tilde{\bm \theta}}^H \\
&=\frac{b^2}{a^4}\cdot {\sf CRLB}(\alpha_m)+\frac{1}{a^2}\cdot{\sf CRLB}(\beta_m)\\
&=\frac{2(d_m+1)\sigma_n^2}{a^4\left\vert h\right\vert^2}.
\end{split}
\end{equation}
From the closed-form CRLB in (\ref{CRLBc}), we can conclude that the star
interconnection strategy is also the optimal interconnection for relative
calibration.

\section{Efficient Estimators for Self-Calibration}\label{SecCaliMethod}

Previously, we have derived the theoretical performance bounds for all the unbiased
estimators. In this section, assuming a total budget of $(M-1)$ transmission lines
to interconnect the $M$ antennas at the BS, under AS-1, we put forward
efficient algorithms to obtain the ML estimates of the calibration coefficients
for any effective interconnection strategy implemented by the BS.
In order to compare with our analytical CRLBs in (\ref{CRLBpro3}), we assume
that the RF gains of the reference antenna , i.e. $\alpha_f$ and $\beta_f$,
are given and fixed. In the meantime, the interconnection channel $h$ is
also assumed to be known. Note the interconnection channel $h$ is
time-invariant and can be estimated in advance.

\subsection{Full Calibration}\label{MLEFull}

From the signal model in (\ref{matrixrece2}), the likelihood function of $\bm{Y}$
conditioned on $\bm \alpha$ and $\bm \beta$ can be written as
$$L(\bm{Y}\vert \bm \alpha,\bm \beta)=\ln p(\bm{Y}\vert \bm \alpha,\bm \beta)=
-\|\bm{Y}-h\bm R \mathcal{A} \bm T\|^2_F+\xi,$$
where $\xi$ includes those terms that do not depend on the unknown parameters.
The ML estimates of the calibration coefficients ${\{\alpha_m,\beta_m\}}_{m=1}^M\setminus\{\alpha_{f},\beta_{f}\}$ can be
obtained by solving the following bi-convex optimization problem:
\begin{equation}\label{MLEpro2}
\begin{split}
[\hat{\bm\alpha},\hat{\bm \beta}]&=\arg \max_{\bm \alpha,\bm \beta}
L(\bm{Y}\vert \bm \alpha,\bm \beta)\\
&=\arg \min_{\bm \alpha,\bm \beta} \left\Vert\bm{Y}-h\bm R \mathcal{A}
\bm T\right\Vert^2_F.
\end{split}
\end{equation}

We can exploit the proposed recursive algorithm as Algorithm \ref{RecursiveFull}
to recover the variables $\bm \alpha$ and $\bm \beta$.
In this algorithm, we put those $s$ antennas whose calibration paths to the reference
antenna include $r$ antennas in addition to the reference antenna and itself
into the set $\mathcal{U}_r=\{r_1,r_2,\ldots,r_s\}$. Specifically, following
the definition of $d_m$ in Proposition 4, we see $d_{r_n}=r, \forall n\in[1,s]$.
Also the variable $p_n$ in Algorithm \ref{RecursiveFull} denotes the index of one
antenna which is in the calibration path of antenna-$r_n$ and interconnected to the
antenna-$r_n$ directly. In Appendix \ref{AppendixFull},
we show that Algorithm \ref{RecursiveFull} can achieve the optimal solution of the
problem in (\ref{MLEpro2}).

\begin{algorithm}[t]
\caption{Recursive Algorithm for Full Calibration with $(M-1)$ Transmission Lines}\label{RecursiveFull}
\begin{algorithmic}[1]
\item {\bf Initialize} $r=0$, $\hat \alpha_f=\alpha_f$, $\hat \beta_f=\beta_f$, and $d_{\max}=\max\{d_1, \ldots, d_{f-1}, d_{f+1}, \ldots, d_M\}$;
\item {\bf While} $r\leq d_{\max}$
\item \quad $s=\vert\mathcal{U}_r\vert$, $\mathcal{U}_r=\{r_1,r_2,\ldots,r_s\}$, $n=1$;
\item \quad {\bf While} $n\le s$
\item \quad \quad $\hat\alpha_{r_n}=\frac{y_{p_n,r_n}}{h \hat\beta_{p_n}}$;
\item \quad \quad $\hat\beta_{r_n}=\frac{y_{r_n,p_n}}{h \hat\alpha_{p_n}}$;
\item \quad \quad $n=n+1$;
\item \quad {\bf End}
\item \quad $r=r+1$;
\item {\bf End}
\end{algorithmic}
\end{algorithm}

For two special interconnection strategies, we can apply Algorithm \ref{RecursiveFull}
and obtain the following ML estimates in closed form.
\begin{itemize}
\item {\it Star Interconnection}:
With calibration measurements in (\ref{scalarrece}),
the solution of (\ref{MLEpro2}) is given by
\begin{equation}\label{Eq:StarSol}
\begin{split}
    \hat{\alpha}_m=\frac{y_{f,m}}{h\beta_f},\
    \hat{\beta}_m=\frac{y_{m,f}}{h\alpha_f},
\end{split}
\end{equation}
where $m\in \{1, 2, \ldots, M\}\setminus \{f\}$.
Note the corresponding mean square errors (MSEs) of the estimates in
(\ref{Eq:StarSol}) are equal to the CRLB results in (\ref{CRLBpro3}), i.e.
\begin{equation}
\begin{split}
    {\sf MSE}(\alpha_m):={\sf E}\left[(\hat{\alpha}_m-\alpha_m)^2\right]=\frac{\sigma_n^2}{b^2\vert h\vert^2},\\
    {\sf MSE}(\beta_m):={\sf E}\left[(\hat{\beta}_m-\beta_m)^2\right]=\frac{\sigma_n^2}{a^2\vert h\vert^2};
\end{split}
\end{equation}

\item {\it Daisy Chain Interconnection}:
With this interconnection strategy, the ML estimates can be derived recursively as
\begin{equation}
\begin{split}
    \hat{\alpha}_m&=\left\{
    \begin{array}{cc}
      \frac{y_{m+1,m}}{h\hat{\beta}_{m+1}} & m<f\\
      \frac{y_{m-1,m}}{h\hat{\beta}_{m-1}} & m>f
    \end{array}
    \right.,\\
    \hat{\beta}_m&=\left\{
    \begin{array}{cc}
      \frac{y_{m,m+1}}{h\hat{\alpha}_{m+1}} & m<f\\
      \frac{y_{m,m-1}}{h\hat{\alpha}_{m-1}} & m>f
    \end{array}
    \right.,
\end{split}
\end{equation}
where $\hat{\alpha}_f=\alpha_f$ and $\hat{\beta}_f=\beta_f$.
\end{itemize}

\subsection{Relative Calibration}\label{MLERelative}

We can rewrite the signal model in (\ref{matrixrece}) as
\begin{equation}
\begin{split}
  \bm Y&=\bm C \bm T\bm H\bm T+\bm N,\\
  &=\bm C\bm \Psi+\bm N,
\end{split}
\end{equation}
where $\bm C:={\sf Diag}\left\{c_1,c_2,\ldots, c_M\right\}$,
$c_m=\beta_m/\alpha_m$, and $\bm \Psi:=\bm T\bm H\bm T$.
Similar to the full calibration in Section \ref{MLEFull},
the ML estimates for the relative calibration coefficients $\{c_m\}_{m=1}^M
\setminus \{c_f\}$ can be obtained  by solving the following optimization
problem \cite{Vieira2017Proposal}:
\begin{equation}\label{MLEproc}
[\hat{\bm c},\hat{\bm \Psi}]=\arg \min_{\bm c, \bm \Psi} \|\bm{Y}-\bm C\bm \Psi\|^2_F.
\end{equation}
With the same notations as in Algorithm \ref{RecursiveFull} for full calibration, we can exploit
Algorithm \ref{RecursiveRela} to derive the relative calibration coefficients.
In Appendix \ref{AppendixRela}, we also show that Algorithm \ref{RecursiveRela} indeed
gives the optimal solution of (\ref{MLEproc}).
In Algorithm \ref{RecursiveRela}, we can see we do not require all
the calibration channels to be the same for relative calibration.
Moreover, we do not need to know the exact values of $h_{p,q}$.
This feature has been exploited in a lot of previous research works, e.g.
\cite{Rogalin2014,wei2016twc,Vieira2017Proposal,BenzinInter2017,Kalten10futnet}.

\begin{algorithm}[t]
\caption{Recursive Algorithm for Relative Calibration}\label{RecursiveRela}
\begin{algorithmic}[1]
\item {\bf Initialize} $r=0$, $\hat c_f=c_f$ and $d_{\max}=\max\{d_1, \ldots, d_{f-1}, d_{f+1}, \ldots, d_M\}$;
\item {\bf While} $r<d_{\max}$
\item \quad $s=\vert\mathcal{U}_r\vert$, $\mathcal{U}_r=\{r_1,r_2,\ldots,r_s\}, n=1;$
\item \quad {\bf While} $n\le s$
\item \quad \quad $\hat{c}_{r_n}=\frac{y_{r_n,p_n}}{y_{p_n,r_n}}\hat c_{p_n}$;
\item \quad \quad $n=n+1$;
\item \quad {\bf End}
\item \quad $r=r+1$;
\item {\bf End}
\end{algorithmic}
\end{algorithm}

For two special interconnection strategies, by applying Algorithm \ref{RecursiveRela},
we have the following closed-form results.
\begin{itemize}
\item {\it Star Interconnection}: For star interconnection, the solution to
(\ref{MLEproc}) is given by
\begin{equation}
  \hat{c}_m=\frac{y_{m,f}}{y_{f,m}}c_f,
\end{equation}
where $m\in\{1, 2, \ldots, M\}\setminus\{f\}$;

\item {\it Daisy Chain Interconnection}:
For daisy chain interconnection, the relative calibration coefficients
can be estimated recursively as
\begin{equation}
  \begin{split}
    \hat{c}_m=\left\{
    \begin{array}{cc}
      \frac{y_{m,m+1}}{y_{m+1,m}}\hat{c}_{m+1} & m<f\\
      \frac{y_{m,m-1}}{y_{m-1,m}}\hat{c}_{m-1} & m>f
    \end{array}
    \right.,
  \end{split}
\end{equation}
where $\hat{c}_f=c_f$.
\end{itemize}

\subsection{More Comments}

In \cite{BenzinInter2017}, the authors have also exploited similar recursive
algorithms to acquire the calibration coefficients.
But they only considered the star interconnection and the daisy interconnection.
On the other hand, our proposed recursive algorithms are more general
and can efficiently solve the ML problem for an arbitrary effective
interconnection strategy.
Besides, we have shown that the recursive algorithms indeed generate the
ML estimates.

\section{Numerical Results}\label{Secnum}

In this section, we provide numerical results to verify our analytical results.
The indices of the antennas are divided into three sets, i.e.
$\mathcal{D}_1:=\{1,\ldots,f-z-1\}$,
$\mathcal{D}_2:=\{f-z, \ldots, f+z\}$, and
$\mathcal{D}_3:=\{f+z+1,\ldots,M\}$.
We define the ``combined interconnection strategy'' as the interconnection strategy
where the antennas in $\mathcal{D}_2$ utilize the daisy chain interconnection
and the antennas in $\mathcal{D}_1$ and $\mathcal{D}_3$ are interconnected to
the $(f-z)$-th and $(f+z)$-th antennas respectively.
In our simulations, we compare the star interconnection, the combined interconnection,
and the daisy chain interconnection for self-calibration at the BS.

Some key parameters assumed in the simulations are listed as follows.
\begin{itemize}
  \item The number of antennas at the BS is set to $M=128$;
  \item The amplitudes of transmit and receive RF gains
  $\left\{\alpha_m,\beta_m\right\}_{m=1}^M$ are equal to $1$,
  and the phases of RF gains are uniformly distributed within $[-\pi, \pi]$;
  \item The transmitted sounding signal is equal to $1$;
  \item The reference antenna is the $65$-th antenna, i.e. $f=65$;
  \item The SNR in the calibration measurements varies from $10$dB to $40$dB;
  \item The value of $z$ for the combined interconnection is set to $5$.
\end{itemize}

\begin{figure*}
\centering
\subfigure[$\sigma_h^2=0$.]
{
    \includegraphics[width=0.45\linewidth]{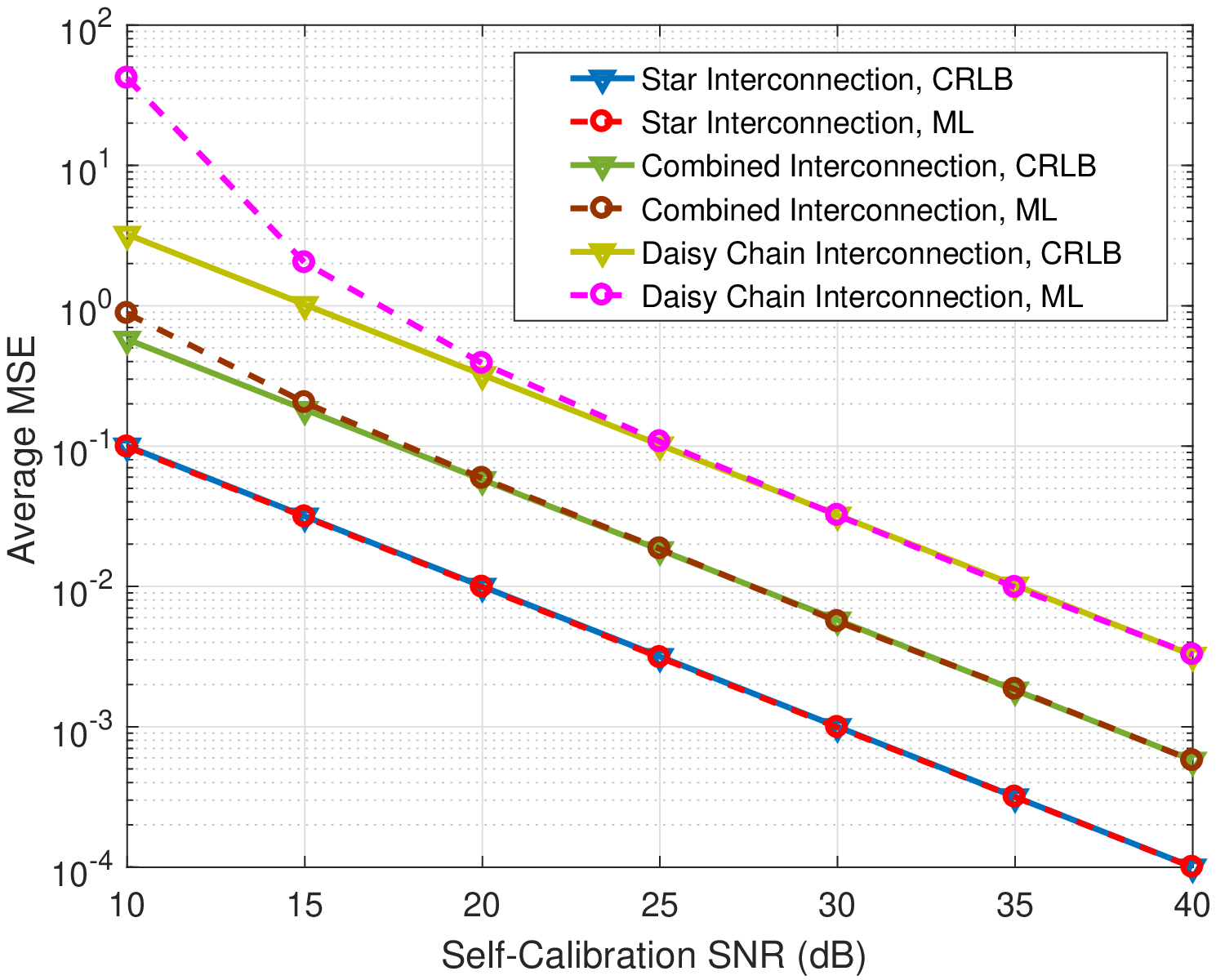}
    \label{fig:MLE_a1}
}
\subfigure[$\sigma_h^2=0.001$.]
{
    \includegraphics[width=0.45\linewidth]{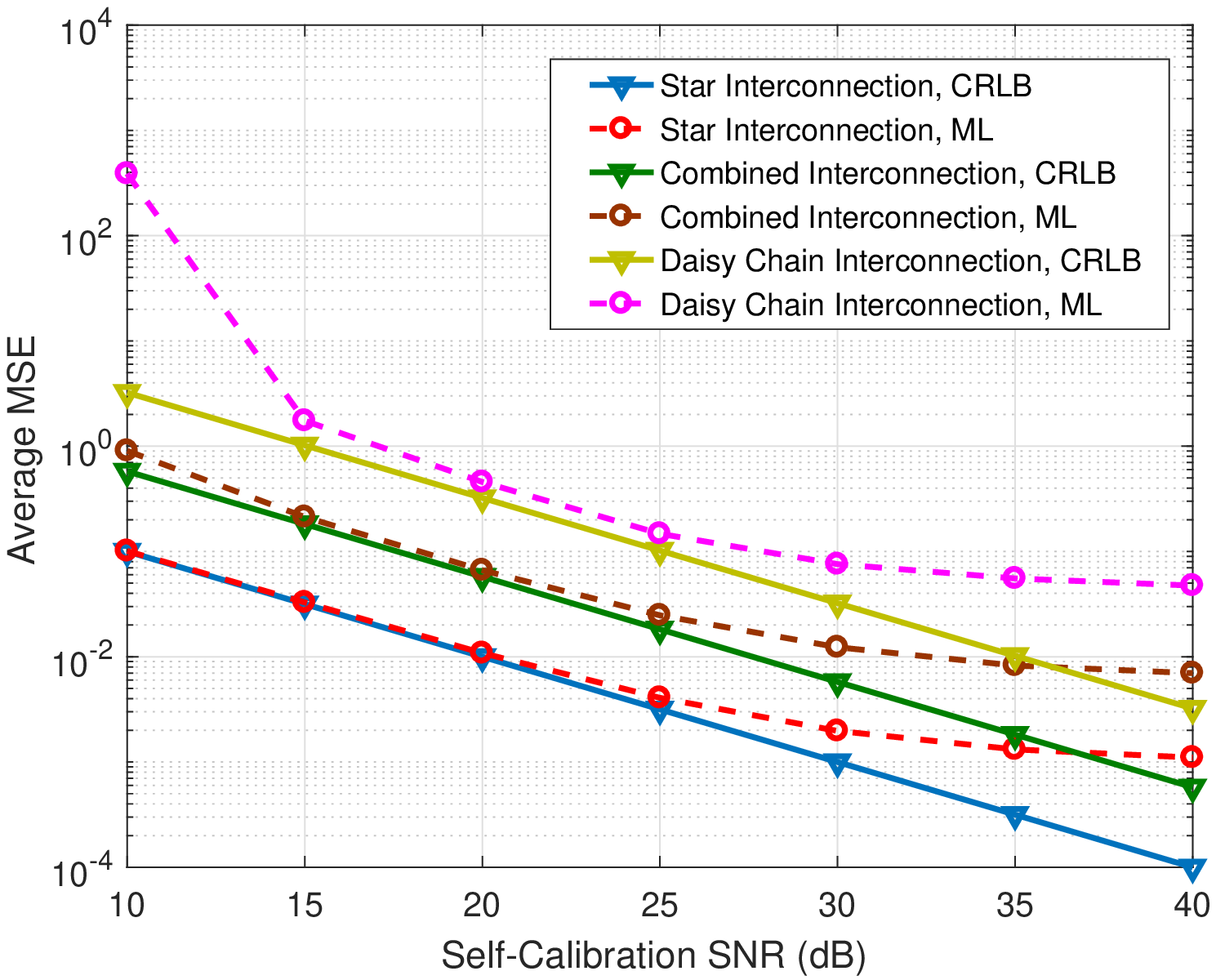}
    \label{fig:MLE_a2}
}
\caption{Full calibration for different interconnection strategies.
(``Star Interconnection'': star interconnection strategy is used for full
calibration at the BS;
``Combined Interconnection'': combined interconnection strategy is used for
full calibration at the BS;
``Daisy Chain Interconnection'': daisy chain interconnection is used for
full calibration at the BS;
``CRLB'': average CRLB over all the unknown calibration coefficients;
{``ML'': simulated average MSE of all the estimated calibration coefficients
for different interconnection strategies with the proposed ML estimators
in Section \ref{SecCaliMethod}.})}
%\vspace{-0.4cm}
\end{figure*}

In practice, we can not ensure all the transmission lines are exactly the same and
all the calibration channels are identical. To take into account this factor,
we assume that $h_{p,q}=h+\tilde{h}_{p,q}$ and $\tilde{h}_{p,q}\sim
\mathcal{CN}(0,\sigma_h^2)$ represents the Gaussian distortions to the calibration
channels. In the ideal case, we have $\sigma_h^2=0$.

In the case of full calibration, Fig. \ref{fig:MLE_a1} shows that the star
interconnection outperforms the other interconnection strategies.
Meanwhile, the proposed recursive algorithm can indeed achieve the
CRLB. In Fig. \ref{fig:MLE_a2}, we set $\sigma_h^2=0.001$ and the simulation
results show that the imperfectness in the designs leads to performance
degradation. However, we see the star interconnection is still the optimal
interconnection strategy.

\begin{figure*}
\centering
\subfigure[$\sigma_h^2=0$.]
{
    \includegraphics[width=0.45\linewidth]{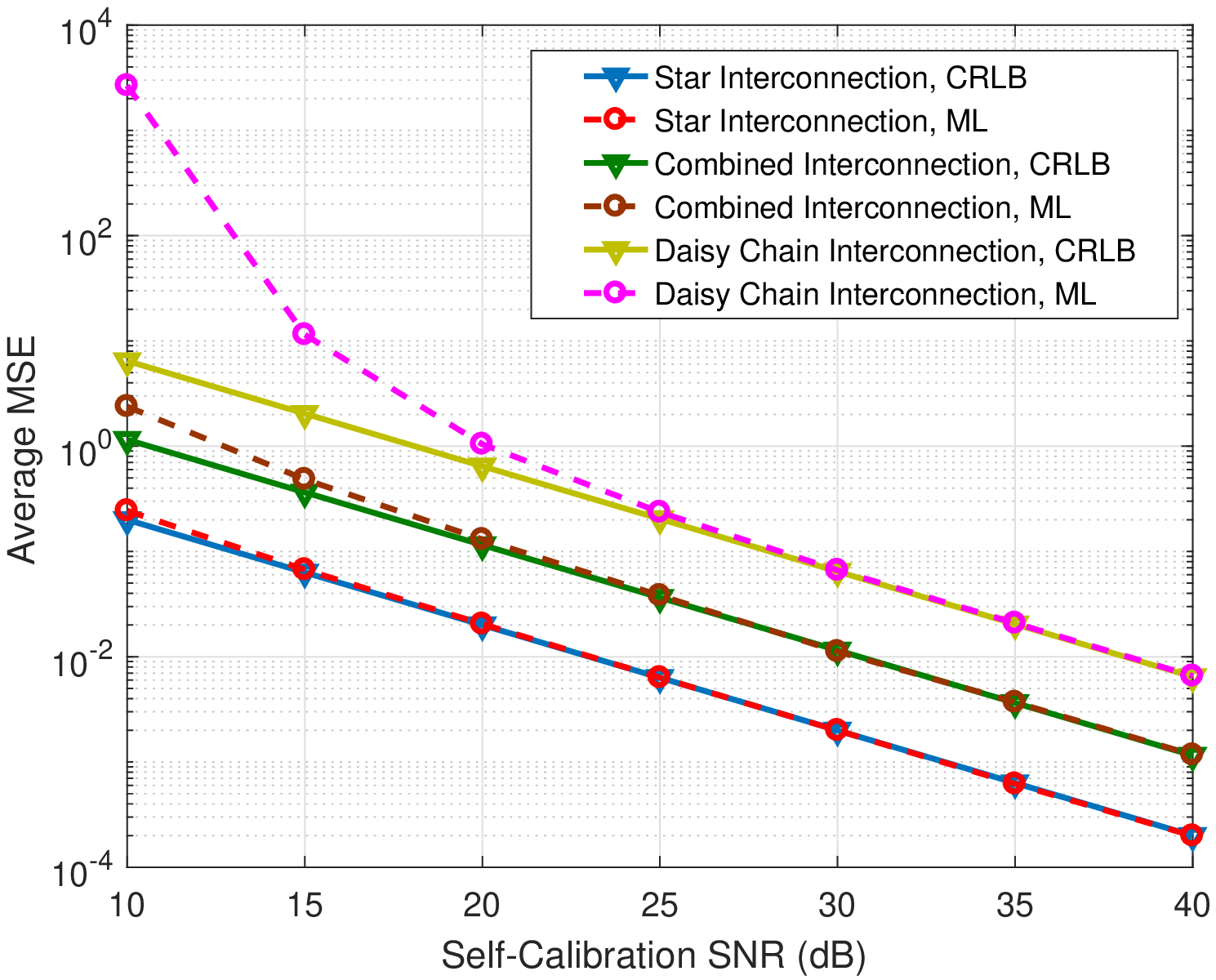}
    \label{fig:MLE_c1}
}
\subfigure[$\sigma_h^2=0.001$.]
{
    \includegraphics[width=0.45\linewidth]{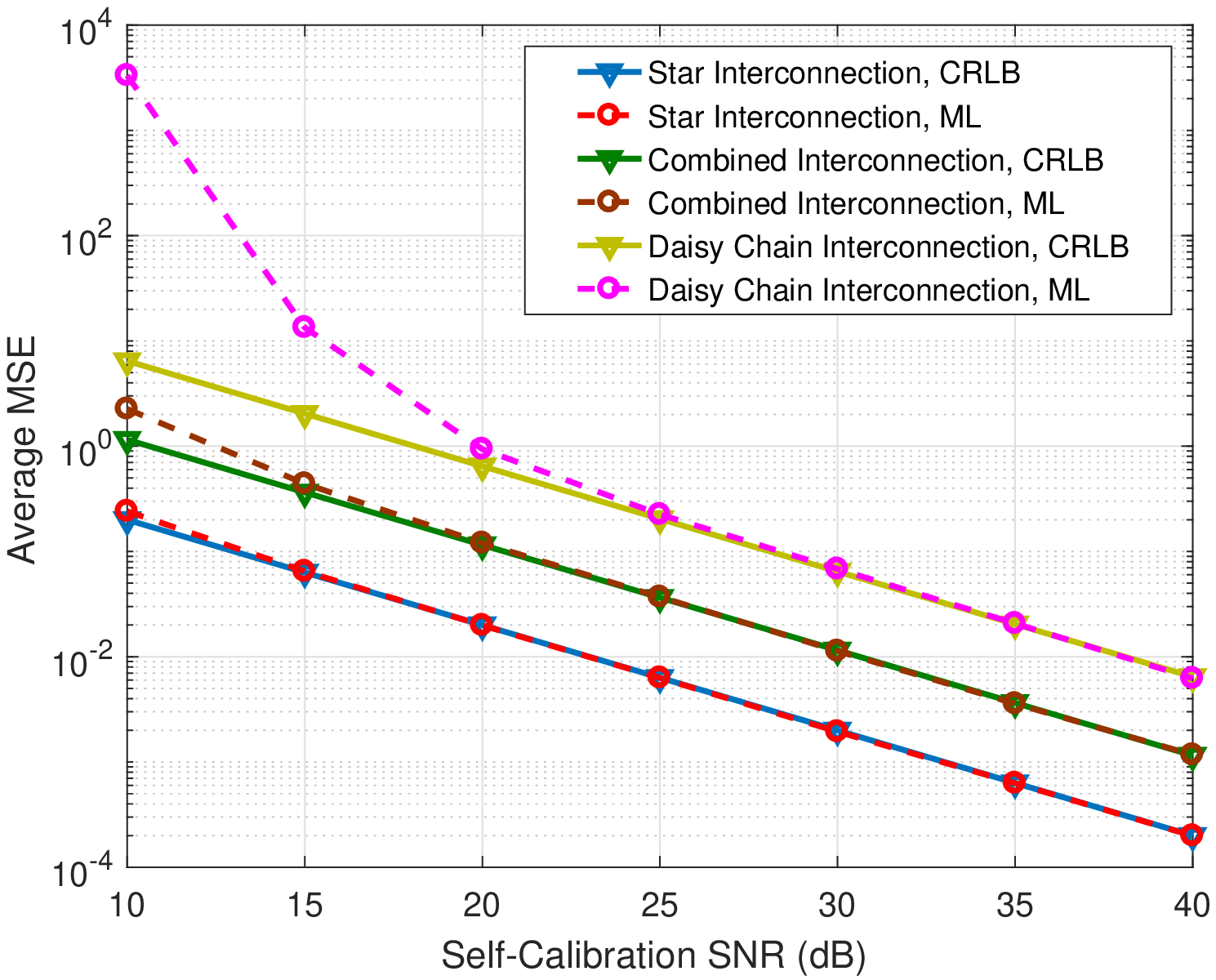}
    \label{fig:MLE_c2}
}
\caption{Relative calibration for different interconnection strategies.
(``Star Interconnection'': star interconnection strategy is used for relative
calibration at the BS;
``Combined Interconnection'': combined interconnection strategy is used
for relative calibration at the BS;
``Daisy Chain Interconnection'': daisy chain interconnection is used
for relative calibration at the BS;
``CRLB'': average CRLB of all the relative calibration coefficients;
``ML'': simulated average MSE of all the estimated relative calibration
coefficients with the ML estimators in Section \ref{SecCaliMethod}.)}
%\vspace{-0.4cm}
\end{figure*}

Fig. \ref{fig:MLE_c1} and Fig. \ref{fig:MLE_c2} show the relative calibration
performance. The simulation results also demonstrate the star interconnection
is the best interconnection strategy. Note that, in the case of relative
calibration, the distortions in the calibration channels only slightly affect
the estimation performance.

\begin{figure*}
\centering
\subfigure[MF precoding.]
{
    \includegraphics[width=0.45\linewidth]{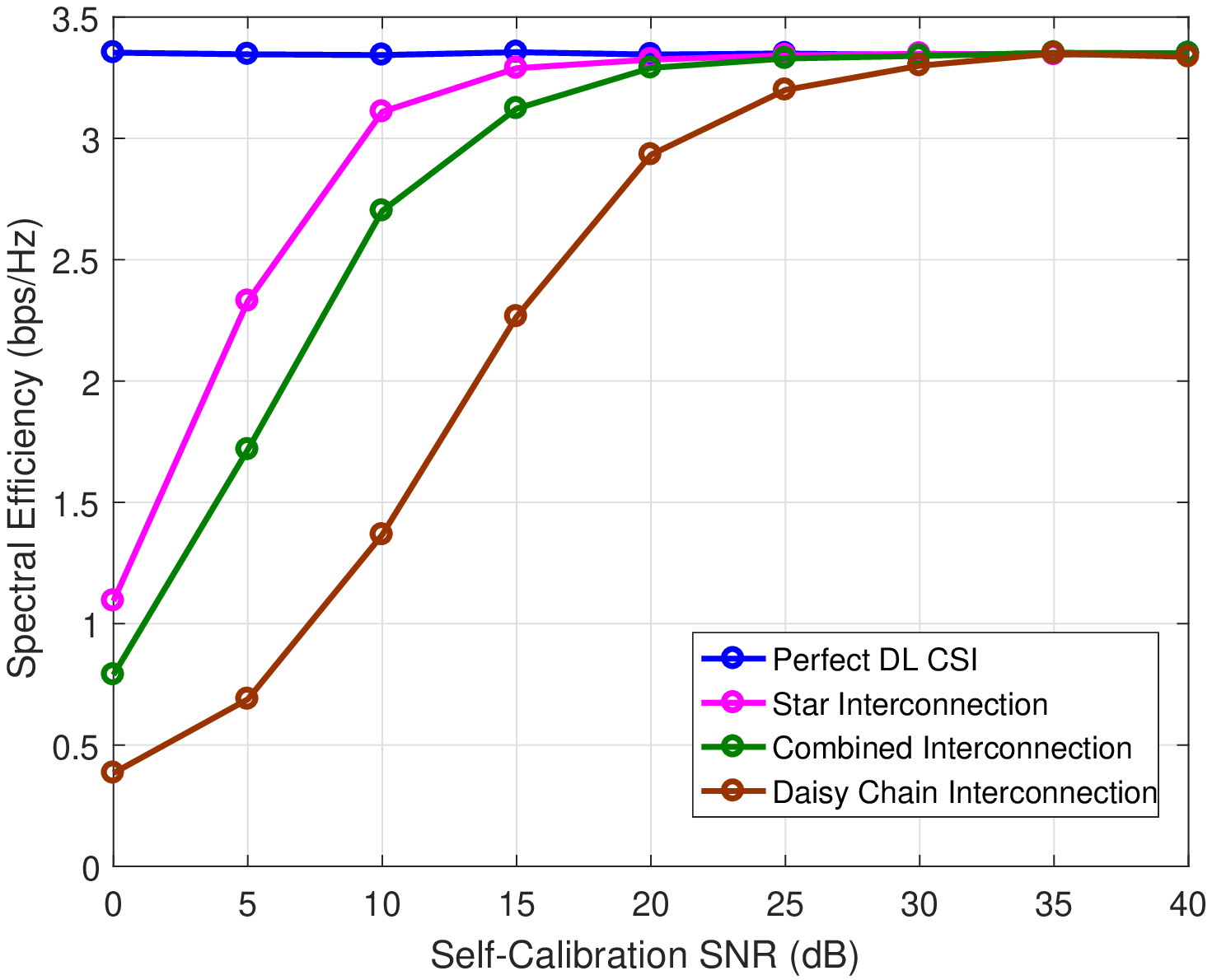}
    \label{fig:DLCapMF}
}
\subfigure[ZF precoding.]
{
    \includegraphics[width=0.45\linewidth]{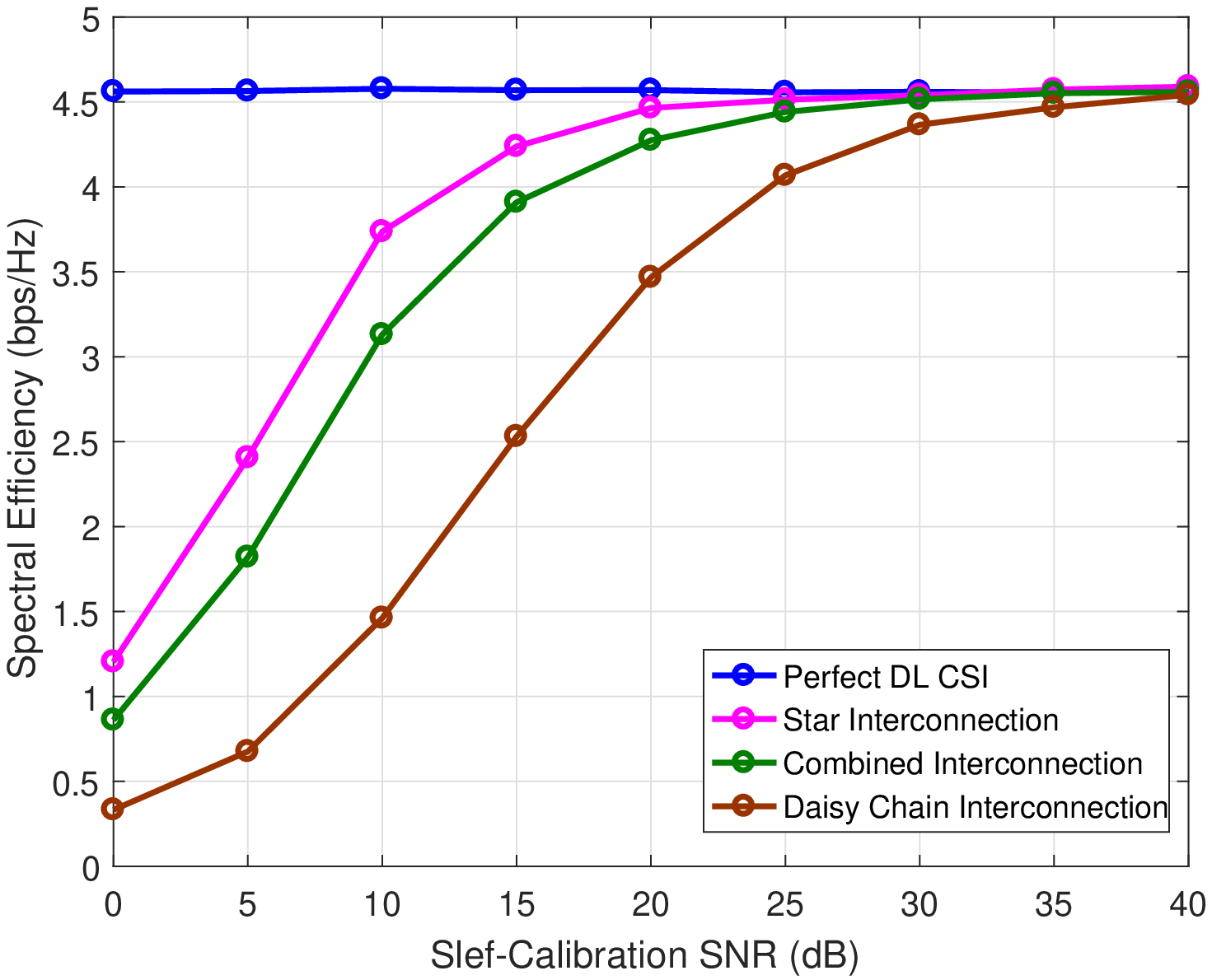}
    \label{fig:DLCapZF}
}
\caption{Average DL spectral efficiency for different interconnection strategies with
full calibration.
(``Star Interconnection'': simulated average DL spectral efficiency when the star
interconnection strategy is used for full calibration;
``Combined Interconnection'': simulated average DL spectral efficiency when
the combined interconnection strategy is used for full calibration;
``Daisy Chain'': simulated average DL spectral efficiency when the daisy chain
interconnection strategy is used for full calibration;
``Perfect DL CSI'': simulated average DL spectral efficiency when the BS has available
perfect DL CSI.)}
\vspace{-0.4cm}
\end{figure*}

In order to see the effects of different interconnection strategies on the DL
spectral efficiency of a massive MIMO system, we simulate a system with one
BS serving $K=6$ MSs. In (\ref{DLsignal}), we set the DL receiver noise variance
to $1$. The DL propagation channels from the BS to the MSs are assumed to be i.i.d.
complex Gaussian with unit variance, i.e. $[H_{PHY}]_{k,m}\sim \mathcal{CN}(0, 1)$.
Assume that the BS carries out either match filter (MF) or zero-forcing (ZF) for DL beamforming \cite{lu14jstsp}. Specifically, we can formulate the precoded data vector
in (\ref{DLsignal}) as $\bm s_D=\frac{1}{\sqrt{\gamma}}\bm W \bm x_D$,
where $\bm W$ contains the beamforming vectors to all $K$ MSs, $\bm x_D$ is a $K\times 1$ signal vector containing the transmitted symbols to each MS, and the scaling factor $\gamma$ normalizes the total transmission power to 1. Assuming $\bm x_D$ is zero mean and satisfies ${\sf E}[\bm x_D\bm x_D^H]=\bm I_K$, we have $\gamma={\sf Tr}\{\bm W^H\bm W\}$. Note for the MF or ZF beamforming, the corresponding precoding matrices are formed as
$\bm W_{MF}=\hat{\bm H}_{DL}^H$ and
$\bm W_{ZF}=\hat{\bm H}_{DL}^H(\hat{\bm H}_{DL}\hat{\bm H}_{DL}^H)^{-1}$ respectively,
where $\hat{\bm H}_{DL}$ denotes the estimate of the DL channel.
Assuming the BS has perfect knowledge about the UL CSIs of all the MSs, by exploiting the TDD channel reciprocity, the estimate of the DL channel is $\hat{\bm H}_{DL}=\bm H_{UL}^T$. In Fig. \ref{fig:DLCapMF} and  Fig. \ref{fig:DLCapZF}, we
simulate the average DL spectral efficiencies when different calibration interconnection
strategies are implemented at the BS. The results show that the star interconnection
achieves the optimal DL spectral efficiency for both MF and ZF precoding.
Meanwhile, the daisy chain interconnection strategy gives the worst DL spectral efficiency
performance.

\section{Conclusions}\label{SecConc}

In this paper, we have studied the interconnect strategies for internal
self-calibration of the large scale antenna array at the BS.
We have derived the CRLBs in estimating the unknown calibration coefficients
for an arbitrary interconnect strategy. Furthermore, closed-form expressions
were derived for each effective interconnection strategy with $(M-1)$
transmission lines. Basing on the theoretical analyses, we have proved
that the star interconnection is the optimal strategy to interconnect the
antennas at the BS for internal self-calibration. Additionally,
we have also put forward efficient recursive algorithms to compute
the ML estimates of those unknown calibration coefficients.
Our results in this paper offer system designers a baseline
philosophy to choose appropriate interconnection strategy for
self-calibration at the BS.

Note in this paper, we have assumed all the transmission lines are of
the same length and gain and our focus is on the optimal interconnection
strategy. In our future works, we can relax these assumptions by
allowing transmission lines of different lengths and seek the most
economic way to interconnect all the antennas.

\appendices
\section{CRLBs for Complex Parameters}\label{AppendixFIM}

By expressing the complex numbers $\alpha_m$ and $\beta_m$ in the form
of real and imaginary parts, i.e. $\alpha_m=\alpha_m^R+\jmath\alpha_m^I$ and
$\beta_m=\beta_m^R+\jmath \beta_m^I$, we can also define the following
$4(M-1)$-by-$1$ vector $\bm \theta$ as
\begin{equation}\label{theta}
  \bm \theta:=\left[\bm \alpha_R^T, \bm \beta_R^T, \bm \alpha_I^T, \bm \beta_I^T \right]^T,
\end{equation}
where
\begin{equation}
\begin{split}
\bm \alpha_R&=\left[\alpha_1^R,\ldots,\alpha_{f-1}^R,\alpha_{f+1}^R,
\ldots, \alpha_M^R\right]^T,\\
\bm \alpha_R&=\left[\beta_1^R, \ldots,\beta_{f-1}^R,  \beta_{f+1}^R,
\ldots, \beta_M^R\right]^T,\\
\bm \alpha_I&=\left[\alpha_1^I,\ldots,\alpha_{f-1}^I,\alpha_{f+1}^I,
\ldots, \alpha_M^I\right]^T,\\
\bm \beta_I&=\left[\beta_1^I, \ldots,\beta_{f-1}^I,  \beta_{f+1}^I,
\ldots, \beta_M^I\right]^T.
\end{split}
\end{equation}

Define the Fisher information matrix of $\bm\theta$ as $\bm J(\bm \theta)$.
The $(i,j)$-th entry of the matrix $\bm J(\bm \theta)$ can be obtained as
\cite{Kay1993}
\begin{equation}\label{FIMoriginal}
  \begin{split}
    \left[\bm J(\bm{\theta})\right]_{i,j}=&{\sf E}\left[\frac{\partial \ln p(\bm y\vert\bm\theta)}{\partial \theta_i}\frac{\partial \ln p(\bm y\vert\bm\theta)}{\partial \theta_j}\right]\\
      =& \text{Tr}\left\{\bm \Sigma^{-1}\frac{\partial \bm \Sigma}{\partial \theta_i}\bm \Sigma^{-1}\frac{\partial \bm \Sigma}{\partial \theta_j}\right\}+
      2 \Re \left\{\frac{\partial \bm{\mu}^H}{\partial \theta_i}\bm \Sigma^{-1}\frac{\partial \bm{\mu}}{\partial \theta_j}\right\}.
  \end{split}
\end{equation}
where $p(\bm y\vert\bm\theta)$ is the complex Gaussian PDF in (\ref{gaussianpdf}).
Since $\bm{\theta}$ is not involved in $\bm \Sigma$, (\ref{FIMoriginal}) can be reduced as
\begin{equation}
  \left[\bm J(\bm{\theta})\right]_{i,j}=\frac{2}{\sigma_n^2} \Re \left\{\frac{\partial \bm{\mu}^H}{\partial \theta_i}\frac{\partial \bm{\mu}}{\partial \theta_j}\right\}.
\end{equation}
Meanwhile, let $\bm \theta_R=[\bm \alpha_R^T,\bm \beta_R^T]^T$ and
$\bm \theta_I=[\bm \alpha_I^T, \bm \beta_I^T]^T$, then the Fisher information
matrix $\bm J(\bm \theta)$ can be rewritten as a block matrix:
\begin{equation}
\bm J(\bm{\theta})=\left[\begin{array}{cc}
\bm J(\bm{\theta}_R)&\bm J(\bm{\theta}_R,\bm{\theta}_I)\\
\bm J(\bm{\theta}_I,\bm{\theta}_R)&\bm J(\bm{\theta}_I)
\end{array}\right].
\end{equation}
Note that
\begin{equation}
\begin{split}
\bm J(\bm{\theta}_R)&={\sf E}\left[\frac{\partial \ln p(\bm y\vert\bm\theta)}{\partial \bm{\theta}_R}\frac{\partial \ln p(\bm y\vert\bm\theta)}{\partial \bm{\theta}_R}\right]\\
&=\frac{2\vert h\vert^{2}}{\sigma_n^2}\cdot\left[\begin{array}{cc}
\bm A& \Re\{\bm D^H\}\\
\Re\{\bm D\}&\bm B
\end{array}\right],\\
\bm J(\bm{\theta}_I,\bm{\theta}_R)&={\sf E}\left[\frac{\partial \ln p(\bm y\vert\bm\theta)}{\partial \bm{\theta}_I}\frac{\partial \ln p(\bm y\vert\bm\theta)}{\partial \bm{\theta}_R}\right]\\
&=\frac{2\vert h\vert^{2}}{\sigma_n^2}\cdot\left[\begin{array}{cc}
\bm O& \Im\{\bm D^H\}\\
\Im\{\bm D\}&\bm O
\end{array}\right],
\end{split}
\end{equation}
where
\begin{equation}
\begin{split}
\bm D&={\sf Diag}\left\{\bm{\beta}\right\}\cdot\bar{\mathcal{A}}\cdot {\sf Diag}\{\bm{\alpha^H}\},\\
  \bm A&={\sf Diag}\left\{\sum\limits_{i\in\mathcal{C}_1}|\beta_i|^2, \ldots, \sum\limits_{i\in\mathcal{C}_m, m\ne f}|\beta_i|^2, \ldots, \sum\limits_{i\in\mathcal{C}_M}|\beta_i|^2\right\},\\
  \bm B&={\sf Diag}\left\{\sum\limits_{i\in\mathcal{C}_1}|\alpha_i|^2, \ldots, \sum\limits_{i\in\mathcal{C}_m, m\ne f}|\alpha_i|^2, \ldots, \sum\limits_{i\in\mathcal{C}_M}|\alpha_i|^2\right\},
\end{split}
\end{equation}
and $\mathcal{C}_m$ denotes the set of the indices of the antennas that are interconnected to the $m$-th antenna directly. Accordingly, the Fisher information matrix of complex parameter $\tilde{\bm \theta}$ in (\ref{Ctheta}) is obtained as\cite{Kay1993}
\begin{equation}\label{FishCtheta}
\begin{split}
\bm J(\tilde{\bm{\theta}})&=\frac{1}{2}\left[\bm J(\bm{\theta}_R)+
\jmath \bm J(\bm{\theta}_I,\bm{\theta}_R)\right]\\
    &=\frac{\vert h\vert^{2}}{\sigma_n^2}\cdot
    \left[\begin{array}{cc}
    \bm A& \bm D^H\\
    \bm D&\bm B
    \end{array}\right].
\end{split}
\end{equation}
Accordingly, we can obtain the CRLB matrix
for $\tilde{\bm \theta}$ with the interconnection strategy $\mathcal{A}$ as
\begin{equation}
{\sf CRLB}(\tilde{\bm{\theta}}|\mathcal{A})=
  \left(\bm J(\tilde{\bm{\theta}})\right)^{-1}.
\end{equation}

\section{Proof of Proposition \ref{pro2}}\label{AppendixPro}
\begin{proof}
Note that $\sum_{m=1}^M t_m=2(M-1)$ since only $(M-1)$ transmission lines are provided.
Let $\mathcal{V}$ denote the set of the indices of the ordinary antennas that are only interconnected to the reference antenna and $\mathcal{W}$ denote the set of the indices of the rest ordinary antennas respectively. Denote that $\vert\mathcal{V}\vert=V$ and $\vert\mathcal{W}\vert=M-V-1$. Obviously, we have $t_v=1, \forall v\in \mathcal{V}$ and $\sum_{v\in\mathcal{V}}t_v=V$.

Except for the star interconnection where all the ordinary antennas
are interconnected to the reference antenna, we have $V<M-1$. Note that for one particular effective interconnection strategy, at least one of the $(M-V-1)$ ordinary antennas, e.g. the $m$-th antenna,
$m\in \mathcal{W}$,  must be interconnected to the reference antenna. Thus we must have $t_f\geq V+1$. Furthermore, we can obtain that
\begin{equation}\label{ineqn}
  \sum_{m\in\mathcal{W}}t_m= 2(M-1)-\sum_{v\in\mathcal{V}}t_v-t_f\leq 2M-2V-3.
\end{equation}

Assuming every ordinary antenna in $\mathcal{W}$ is
interconnected to two or more other antennas, we have
$t_m\geq2, \forall m\in \mathcal{W}$. Then the following
inequality must hold:
\begin{equation}
  \sum_{m\in\mathcal{W}}t_m\geq 2\vert\mathcal{W}\vert=2M-2V-2,
\end{equation}
which is in contradiction to (\ref{ineqn}).
Thus, from the definition of the set $\mathcal{W}$,
we can conclude that there must exist one ordinary antenna
in $\mathcal{W}$ which is only interconnected to another
ordinary antenna.
\end{proof}

\section{Proof of Proposition \ref{pro3new}}\label{AppendiaxFimupdate}
\begin{proof}
Firstly, we consider an interconnection strategy $\mathcal{A}^{(k)}$, where $1\leq W\leq M-2$ ordinary antennas are not interconnected to the reference antenna.
According to Proposition \ref{pro2}, in $\mathcal{A}^{(k)}$, we can find one ordinary antenna, i.e. the $n_k$-th antenna, which is only connected to another ordinary antenna, i.e.
the $u_k$-th antenna.
By breaking the connection to the $u_k$-th antenna and interconnecting the $n_k$-th antenna to the reference antenna, we can obtain an updated interconnection strategy $\mathcal{A}^{(k+1)}$.
Clearly, only $(W-1)$ ordinary antennas are not interconnected to the reference antenna directly in the strategy $\mathcal{A}^{(k+1)}$.

There are two facts about the interconnection strategies $\mathcal{A}^{(k)}$ and $\mathcal{A}^{(k+1)}$ worth noting.
The first fact is that the Fisher information matrix $\bm J^{(k)}(\tilde{\bm{\theta}})$
only differs from the matrix $\bm J^{(k+1)}(\tilde{\bm{\theta}})$
in six elements.
Specifically, these six elements include two diagonal elements in the positions ($\bar u_k,\bar u_k $) and ($\bar u_k',\bar u_k'$), and four non-diagonal elements in the positions ($\bar n_k,\bar u_k'$), ($\bar u_k,\bar n_k'$), ($\bar u_k',\bar n_k$), and ($\bar n_k',\bar u_k$).
The second fact is that the four non-diagonal elements
of $\bm J^{(k+1)}(\tilde{\bm{\theta}})$
in the rows: $\bar n_k$, $\bar n_k'$ and columns: $\bar n_k$, $\bar n_k'$
are zeros.

Thanks to the special structure of $\bm J^{(k)}(\tilde{\bm{\theta}})$, only
two diagonal elements in the positions ($\bar u_k,\bar u_k $) and ($\bar u_k',\bar u_k'$)
are changed when we apply elementary transformations to the matrix
$\bm J^{(k)}(\tilde{\bm{\theta}})$ to null the aforementioned four non-diagonal
elements.
Specifically, we can carry out the matrix elementary transformations as
$\bm L^{(k)} \bm J^{(k)}(\tilde{\bm{\theta}}) \bm L'^{(k)}$,
where
\begin{eqnarray}
  \bm L^{(k)} := \bm L_{\bar u'_k,\bar n_k}\left(-\frac{\beta_{u} \alpha_{n}^*}{b^2}\right) \bm L_{\bar u_k,\bar n'_k}\left(-\frac{\beta_{n}^* \alpha_{u}}{a^2}\right),\\
  \bm L'^{(k)} := \bm L_{\bar n'_k,\bar u_k}\left(-\frac{\beta_{n} \alpha_{u}^*}{a^2}\right) \bm L_{\bar n_k,\bar u'_k}\left(-\frac{\beta_{u}^* \alpha_{n}}{b^2}\right).
\end{eqnarray}
It can be verified that the matrix $\bm J^{(k)}(\tilde{\bm{\theta}})$ becomes
equal to $\bm J^{(k+1)}(\tilde{\bm{\theta}})$ after the above
elementary transformations, i.e.
$\bm J^{(k+1)}(\tilde{\bm{\theta}})=\bm L^{(k)} \bm J^{(k)}(\tilde{\bm{\theta}}) \bm L'^{(k)}$.
\end{proof}

\section{Proof of Proposition \ref{pro4}}\label{AppendixPro4}
\begin{proof}
Let $d_m$ represent the number of intermediate antennas along the calibration path
between the reference antenna and the $m$-th antenna in the original interconnection
strategy $\mathcal{A}^{(0)}$.
Starting from $\mathcal{A}^{(k)}$, $k\in[0,W-1]$, we will have one
new interconnection strategy $\mathcal{A}^{(k+1)}$ by performing the
$k$-th update in (\ref{Fimtransk}) as described in Proposition 3. Clearly, we have
$(W-k)$ ordinary antennas in $\mathcal{A}^{(k)}$ which are not
interconnected to the reference antenna.
Let antenna-$n_k$ be the identified ordinary antenna that is only interconnected to
the antenna-$u_k$ in the interconnection strategy $\mathcal{A}^{(k)}$.
First, we can obtain the following two intermediate results:
\begin{enumerate}
\item
The relationship between the CRLBs corresponding to the interconnection strategies
$\mathcal{A}^{(k)}$ and $\mathcal{A}^{(k+1)}$ is described by (\ref{CRLBk}).
Due to the special structure of the elementary matrices, the diagonal elements of
$(\bm J^{(k)}(\tilde{\bm{\theta}}))^{-1}$ are the same as those of
$(\bm J^{(k+1)}(\tilde{\bm{\theta}}))^{-1}$ except the $\bar n_k$-th and
the $\bar n'_k$-th diagonal elements which are given by
\begin{equation}\label{res_1}
\begin{split}
  &\left[\left(\bm J^{(k)}(\tilde{\bm{\theta}})\right)^{-1}\right]_{\bar n_k,\bar n_k} = \left[\left(\bm J^{(k+1)}(\tilde{\bm{\theta}})\right)^{-1}\right]_{\bar n_k,\bar n_k} \\
  &+\frac{a^2}{b^2} \cdot  \left[\left(\bm J^{(k+1)}(\tilde{\bm{\theta}})\right)^{-1}\right]_{\bar u'_k,\bar u'_k},\\
  &\left[\left(\bm J^{(k)}(\tilde{\bm{\theta}})\right)^{-1}\right]_{\bar n'_k,\bar n'_k}=\left[\left(\bm J^{(k+1)}(\tilde{\bm{\theta}})\right)^{-1}\right]_{\bar n'_k,\bar n'_k}\\
  &+ \frac{b^2}{a^2} \cdot \left[\left(\bm J^{(k+1)}(\tilde{\bm{\theta}})\right)^{-1}\right]_{\bar u_k,\bar u_k}.
\end{split}
\end{equation}
The notations $\bar n_k$, $\bar n'_k$, $\bar u_k$ and $\bar u'_k$ are defined
as in (\ref{ele_mat}).

\item
From the results in (\ref{res_1}), we can see the
$\bar n_k$-th and $\bar n'_k$-th diagonal elements get
updated only when we derive the CRLB matrix
$\left(\bm J^{(k)}(\tilde{\bm{\theta}})\right)^{-1}$ from
$\left(\bm J^{(k+1)}(\tilde{\bm{\theta}})\right)^{-1}$.
This is due to the fact that $n_{k'}\neq n_{k}$, $\forall k'\neq k$.
Hence, $\forall r\in(0,W-k)$, we see
the $\bar n_k$-th and $\bar n'_k$-th diagonal elements of the CRLB
matrix $\left(\bm J^{(k+r)}(\tilde{\bm{\theta}})\right)^{-1}$
are the same as those in the CRLB matrix
$\left(\bm J^{(W)}(\tilde{\bm{\theta}})\right)^{-1}=\left(\bm J_{\text{star}}(\tilde{\bm{\theta}})\right)^{-1}$.
It then follows that, $\forall r\in(0, W-k)$,
\begin{equation}\label{res2_1}
\begin{split}
    \left[\left(\bm J^{(k+r)}(\tilde{\bm{\theta}})\right)^{-1}\right]_{\bar n_k,\bar n_k}=\frac{\sigma_n^2}{b^2|h|^{2}},\\
  \left[\left(\bm J^{(k+r)}(\tilde{\bm{\theta}})\right)^{-1}\right]_{\bar n_k',\bar n_k'}=\frac{\sigma_n^2}{a^2|h|^{2}}.
\end{split}
\end{equation}
In the mean time, when $0<r\leq k$, we also see the
$\bar n_k$-th and $\bar n'_k$-th diagonal elements of the CRLB matrix
$\left(\bm J^{(k-r)}(\tilde{\bm{\theta}})\right)^{-1}$
are the same as those in
$\left(\bm J^{(k)}(\tilde{\bm{\theta}})\right)^{-1}$.
So we can get, $\forall r\in(0, k]$,
\begin{equation}\label{res2_2}
\begin{split}
   \left[\left(\bm J^{(k-r)}(\tilde{\bm{\theta}})\right)^{-1}\right]_{\bar n_k,\bar n_k}=\left[\left(\bm J^{(k)}(\tilde{\bm{\theta}})\right)^{-1}\right]_{\bar n_k,\bar n_k},\\
   \left[\left(\bm J^{(k-r)}(\tilde{\bm{\theta}})\right)^{-1}\right]_{\bar n_k',\bar n_k'}=\left[\left(\bm J^{(k)}(\tilde{\bm{\theta}})\right)^{-1}\right]_{\bar n_k,\bar n_k}.
\end{split}
\end{equation}
\end{enumerate}

With the above two intermediate results, we can apply the method of
mathematical induction to complete the proof as follows.\\
\noindent$\bullet$
When $d_m=0$, we see
$m\not \in\{n_0,n_1,\cdots,n_{W-1}\}$ and the
$\bar m$-th and $\bar m'$-th diagonal elements of the CRLB matrix $\left(\bm J^{(0)}(\tilde{\bm{\theta}})\right)^{-1}$ are not changed when we obtain
it from the CRLB matrix $\left(\bm J_{\text{star}}(\tilde{\bm{\theta}})\right)^{-1}$.
Note that $\bar m$ and $\bar m'$ denote the indices of the rows corresponding
to $\alpha_m$ and $\beta_m$ in $\tilde{\bm{\theta}}$ respectively.
Thus the CRLBs for the parameters $\alpha_m$ and $\beta_m$ are given by
\begin{equation}\label{CRLB0}
  \begin{split}
    {\sf CRLB}(\alpha_m)=\left[\left(\bm J_{\text{star}}(\tilde{\bm{\theta}})\right)^{-1}\right]_{\bar m,\bar m}=\frac{\sigma_n^2}{b^2|h|^{2}},\\
    {\sf CRLB}(\beta_m)=\left[\left(\bm J_{\text{star}}(\tilde{\bm{\theta}})\right)^{-1}\right]_{\bar m',\bar m'}=\frac{\sigma_n^2}{a^2|h|^{2}}.
  \end{split}
\end{equation}

\noindent$\bullet$
When $d_m=1$, we have $m=n_k$ and $d_{u_k}=0$ for one particular
$k$ in $[0,W-1]$. By applying the results in (\ref{res2_2}), we can further
obtain
\begin{equation}
\label{CRLB1_1}
\left[\left(\bm J^{(0)}(\tilde{\bm{\theta}})\right)^{-1}\right]_{\bar n_k,\bar n_k}=
\left[\left(\bm J^{(k)}(\tilde{\bm{\theta}})\right)^{-1}\right]_{\bar n_k,\bar n_k}.
\end{equation}
With (\ref{res_1}), the above result can be rewritten as
\begin{equation}\label{CRLB1_2}
\begin{split}
 \left[\left(\bm J^{(0)}(\tilde{\bm{\theta}})\right)^{-1}\right]&_{\bar n_k,\bar n_k} =\left[\left(\bm J^{(k+1)}(\tilde{\bm{\theta}})\right)^{-1}\right]_{\bar n_k,\bar n_k}\\
 & + \frac{a^2}{b^2} \cdot\left[\left(\bm J^{(k+1)}(\tilde{\bm{\theta}})\right)^{-1}\right]_{\bar u_k',\bar u_k'},\\
\end{split}
\end{equation}
where
$[(\bm J^{(k+1)}(\tilde{\bm{\theta}}))^{-1}]_{\bar n_k,\bar n_k}=\frac{\sigma_n^2}{b^2|h|^{2}}$ due to (\ref{res2_1}) and
$[(\bm J^{(k+1)}(\tilde{\bm{\theta}}))^{-1}]_{\bar u_k',\bar u_k'}=\frac{\sigma_n^2}{a^2|h|^{2}}$ due to the fact that $d_{u_k}=0$ and
the result in (\ref{CRLB0}).
Then (\ref{CRLB1_2}) can be rewritten as
\begin{equation*}
 {\sf CRLB}(\alpha_m)=\left[\left(\bm J^{(0)}(\tilde{\bm{\theta}})\right)^{-1}\right]_{\bar n_k,\bar n_k} =\frac{2\sigma_n^2}{b^2|h|^{2}}.
\end{equation*}
Similarly, we can obtain the result for $\beta_m$ as
\begin{equation*}
{\sf CRLB}(\beta_m)=
\left[\left(\bm J^{(0)}(\tilde{\bm{\theta}})\right)^{-1}\right]_{\bar n_k',\bar n_k'} =\frac{2\sigma_n^2}{b^2|h|^{2}}.
\end{equation*}

\noindent$\bullet$
We assume Proposition \ref{pro4} is true for each ordinary antenna with
$s\geq 1$ additional antennas along its calibration path.
For antenna-$m$ with $d_m=s+1$, we can have one particular
$k$ in $[0,W-1]$ and $r\geq 1$ such that
$n_k=m$, $u_k=n_{k+r}$, and $d_{u_k}=d_{n_{k+r}}=d_m-1=s$.
By applying the results in (\ref{res_1})-(\ref{res2_2}),
we can obtain
\begin{equation*}
\begin{split}
& {\sf CRLB}(\alpha_m)=\left[\left(\bm J^{(0)}(\tilde{\bm{\theta}})\right)^{-1}\right]_{\bar n_k,\bar n_k} \\
=&\left[\left(\bm J^{(k+1)}(\tilde{\bm{\theta}})\right)^{-1}\right]_{\bar n_k,\bar n_k} + \frac{a^2}{b^2} \cdot\left[\left(\bm J^{(k+1)}(\tilde{\bm{\theta}})\right)^{-1}\right]_{\bar u_k',\bar u_k'}\\
=&\frac{(d_{m}+1)\sigma_n^2}{b^2|h|^{2}},
\end{split}
\end{equation*}
where we have utilized the fact that
$[(\bm J^{(k+1)}(\tilde{\bm{\theta}}))^{-1}]_{\bar n_k,\bar n_k}=
\frac{\sigma_n^2}{b^2|h|^2}$ due to (\ref{res2_1}) and our starting
assumption. Specifically, due to the result in (\ref{res2_2}), we have
\begin{equation*}
\begin{split}
\left[\left(\bm J^{(k+1)}(\tilde{\bm{\theta}})\right)^{-1}\right]_{\bar u_k',\bar u_k'}&=\left[\left(\bm J^{(k+1)}(\tilde{\bm{\theta}})\right)^{-1}\right]_{\bar n_{k+r}',\bar n_{k+r}'}\\
&=\left[\left(\bm J^{(k+r)}(\tilde{\bm{\theta}})\right)^{-1}\right]_{\bar n_{k+r}',\bar n_{k+r}'}.
\end{split}
\end{equation*}
Since $d_{n_{k+r}}=s$, according to our starting assumption, we have
$[(\bm J^{(k+r)}(\tilde{\bm{\theta}}))^{-1}]_{\bar n_{k+r}',\bar n_{k+r}'}=\frac{(s+1)\sigma_n^2}{a^2|h|^{2}}$.
Similarly, we can obtain the CRLB for $\beta_m$ as
\begin{equation*}
{\sf CRLB}(\beta_m)=\frac{(d_m+1)\sigma_n^2}{a^2|h|^{2}}.
\end{equation*}
Thus Proposition \ref{pro4} is also true for each ordinary antenna with
$(s+1)$ additional antennas along its calibration path.
\end{proof}

\section{ML Estimator for Full Calibration}\label{AppendixFull}

Let $d_{\max}=\max\{d_1, \ldots, d_{f-1}, d_{f+1}, \ldots, d_M\}$. The optimization problem
in (\ref{MLEpro2}) can be rewritten as
\begin{equation}\label{eqfull1}
\begin{split}
[\hat{\bm\alpha},\hat{\bm \beta}]=\arg \min_{\bm \alpha,\bm \beta} \sum_{r=0}^{d_{\max}}\sum_{n=1}^{\vert\mathcal{U}_r\vert}\big(&\vert y_{r_n,p_n}- \beta_{r_n} h \alpha_{p_n}\vert^2\\
&+\vert y_{p_n,r_n}- \beta_{p_n} h \alpha_{r_n}\vert^2\big),
\end{split}
\end{equation}
where $r_n\in\mathcal{U}_r$.

When $r=0$, we have $p_n=f$ for $n=1,2,\ldots,\vert\mathcal{U}_r\vert$.
According to Algorithm \ref{RecursiveFull}, we can
estimate $\alpha_{r_n}$ and $\beta_{r_n}$ as
\begin{equation}\label{eqfull2}
\begin{split}
\hat\alpha_{r_n}&=\frac{y_{f,r_n}}{h \beta_{f}}, \ \ \hat\beta_{r_n}=\frac{y_{r_n,f}}{h \alpha_{f}}.
\end{split}
\end{equation}
Thus, for $r=0$, from (\ref{eqfull1}) and (\ref{eqfull2}), we have
\begin{equation}\label{eqfull3}
\sum_{n=1}^{\vert\mathcal{U}_r\vert}\left(\vert y_{r_n,f}- \hat\beta_{r_n} h
\alpha_{f}\vert^2+\vert y_{f,r_n}- \beta_{f} h \hat\alpha_{r_n}\vert^2\right)=0.
\end{equation}

Similarly, when $r\ge 1$, according to Algorithm \ref{RecursiveFull},
we can estimate $\alpha_{r_n}$ and $\beta_{r_n}$ as
\begin{equation}
\begin{split}
\hat\alpha_{r_n}&=\frac{y_{p_n,r_n}}{h \hat\beta_{p_n}},  \ \ \hat\beta_{r_n}=\frac{y_{r_n,p_n}}{h \hat \alpha_{p_n}}.
\end{split}
\end{equation}
Accordingly, we have the following equality:
\begin{equation}\label{eqfull4}
\sum_{n=1}^{\vert\mathcal{U}_r\vert}\left(\vert y_{r_n,p_n}- \hat\beta_{r_n} h
\hat\alpha_{p_n}\vert^2+\vert y_{p_n,r_n}- \hat\beta_{p_n} h \hat\alpha_{r_n}\vert^2\right)=0.
\end{equation}

From (\ref{eqfull3}), (\ref{eqfull4}), we have $\left\Vert\bm{Y}-\bm R \bm H \bm T\right\Vert^2_F=0$ and the object function in (\ref{MLEpro2}) achieves
the minimum since $\left\Vert\bm{Y}-\bm R \bm H \bm T\right\Vert^2_F\geq0$.

\section{ML Estimator for Relative Calibration}\label{AppendixRela}
Let $d_{\max}=\max\{d_1, \ldots, d_{f-1}, d_{f+1}, \ldots, d_M\}$. The optimization
problem in (\ref{MLEproc}) can be rewritten as
\begin{equation}\label{eqrela1}
\begin{split}
  [\hat{\bm\alpha},\hat{\bm \beta}]=\arg \min_{\bm \alpha,\bm \beta}
\sum_{r=0}^{d_{\max}}\sum_{n=1}^{\vert\mathcal{U}_r\vert}\big(&\vert y_{p_n,r_n}- c_{p_n}
h \bm\Psi_{p_n,r_n}\vert^2 \\
+&\vert y_{r_n,p_n}- c_{r_n} h \bm\Psi_{r_n,p_n}\vert^2\big),
\end{split}
\end{equation}
where $r_n\in\mathcal{U}_r$ and $\bm\Psi_{r_n,p_n}=\bm\Psi_{p_n,r_n}$.

In Algorithm \ref{RecursiveRela}, if $r=0$, we have $p_n=f$ for $n=1,2,\ldots,\vert\mathcal{U}_r\vert$, and
we estimate $c_{r_n}$ and $\bm \Psi_{r_n,f}$ as
\begin{equation}\label{eqrela2}
\begin{split}
&\hat{\bm\Psi}_{r_n,f}=\frac{y_{f,r_n}}{c_f h},  \ \ \hat c_{r_n}=\frac{y_{r_n,f}}{h \hat{\bm\Psi}_{r_n,f}}=\frac{y_{r_n,f}}{y_{f,r_n}}c_f.
\end{split}
\end{equation}
Thus, for $r=0$, from (\ref{eqrela1}) and (\ref{eqrela2}), we have
\begin{equation}\label{eqrela3}
\sum_{n=1}^{\vert\mathcal{U}_r\vert}\big(\vert y_{f,r_n}- c_{f} h
\bm\hat\Psi_{f,r_n}\vert^2+\vert y_{r_n,f}- \hat c_{r_n} h \bm\Psi_{r_n,f}\vert^2\big)=0.
\end{equation}

When $r\geq1$, according to the algorithm \ref{RecursiveRela}, we can estimate
$c_{r_n}$ and $\bm \Psi_{r_n,p_n}$ as
\begin{equation}
\begin{split}
&\hat{\bm\Psi}_{r_n,p_n}=\frac{y_{p_n,r_n}}{\hat c_{p_n} h}, \ \ \hat c_{r_n}=\frac{y_{r_n,p_n}}{h \hat{\bm\Psi}_{r_n,p_n}}=
\frac{y_{r_n,p_n}}{y_{p_n,r_n}}\hat c_{r_{}n}.
\end{split}
\end{equation}
Then we can have the following equality:
\begin{equation}\label{eqrela4}
\sum_{n=1}^{\vert\mathcal{U}_r\vert}\big(\vert y_{p_n,r_n}- \hat c_{p_n} h
\bm\hat\Psi_{p_n,r_n}\vert^2+\vert y_{r_n,p_n}- \hat c_{r_n} h \bm\hat\Psi_{r_n,p_n}\vert^2=0\big).
\end{equation}

From (\ref{eqrela3}) and (\ref{eqrela4}), we have
$\left\Vert\bm{Y}-\bm C \bm \Psi\right\Vert^2_F=0$ and the object function
in (\ref{MLEpro2}) achieves the minimum since
$\left\Vert\bm{Y}-\bm C \bm \Psi\right\Vert^2_F\geq0$.

%\newpage
\bibliographystyle{IEEE}

\end{document}